\documentclass[aps,pre,twocolumn,epsfig,floats]{revtex4-2}
\usepackage{amsmath}
\usepackage{amssymb}
\usepackage{color}
\usepackage{graphicx}
\usepackage{epsfig}
\usepackage{amsfonts}
\usepackage{mathrsfs}
\usepackage{mathtools}
\usepackage{amsmath}
\usepackage{float}
\usepackage{physics}
\usepackage{hyperref}
\hypersetup{
  colorlinks=true,
  citecolor=blue,
  linkcolor=blue,
  urlcolor=blue}

\begin{document}
\title{Non-equilibrium dynamics of the Jaynes-Cummings dimer}

\author{G. Vivek, Debabrata Mondal, and S. Sinha}
\affiliation{Indian Institute of Science Education and
	Research-Kolkata, Mohanpur, Nadia-741246, India}

\begin{abstract}
We investigate the non-equilibrium dynamics of a Josephson coupled Jaynes-Cummings dimer in the presence of Kerr nonlinearity, which can be realized in the cavity and circuit quantum electrodynamics systems. The semiclassical dynamics is analyzed systematically to chart out a variety of photonic Josephson oscillations and their regime of stability.  
Different types of transitions between the dynamical states lead to the self trapping phenomenon, which results in photon population imbalance between the two cavities.
We also study the dynamics quantum mechanically to identify characteristic features of different steady states and to explore fascinating quantum effects, such as spin dephasing, phase fluctuation, and revival phenomena of the photon field, as well as the entanglement of spin qubits. For a particular `self trapped' state, the mutual information between the atomic qubits exhibits a direct correlation with the photon population imbalance, which is promising for generating photon mediated entanglement between two non interacting qubits in a controlled manner. Under a sudden quench from stable to unstable regime, the photon distribution exhibits phase space mixing with a rapid loss of coherence, resembling a thermal state. Finally, we discuss the relevance of the new results in experiments, which can have applications in quantum information processing and quantum technologies.
\end{abstract}

\date{\today}
\maketitle
\section{Introduction}
Recent advancements in cavity and circuit quantum electrodynamics (QED) have paved the way to study the non-equilibrium dynamics in quantum systems \cite{Angelakis,Serge_Haroche,Steven_Girvin}, apart from their potential application to quantum information processing \cite{Serge_Haroche,Steven_Girvin}.  
Moreover, such atom-photon interacting systems exhibit various fascinating phenomena, some of which include quantum phase transition \cite{QPT_0, QPT_Chaos_Brandes, QPT_2}, the onset of chaos \cite{QPT_Chaos_Brandes}, thermalization \cite{Haake_2012,Lea_2023}, and the formation of quantum scars \cite{Scar1,Scar2}, which has attracted significant interest in recent years.
In addition, a range of quantum effects associated with photons can also be explored, for example, the collapse and revival phenomenon \cite{Revival_1,Revival_2,Revival_3}, formation of Schr\"{o}dinger's cat state \cite{Cat_1,Cat_2,Cat_3,Cat_4,Cat_5} and non-classical state of light \cite{Non_classical_1,Non_classical_2,Non_classical_3,Non_classical_4,Non_classical_5,Non_classical_6}.
Current experiments have demonstrated that coupling atomic condensates to the cavity mode can lead to fascinating phenomena like the formation of super solid phase \cite{Esslinger_SS_2,Esslinger_SS_3} and non-equilibrium transition \cite{Hemmarich}.
It is also important to note that photon loss and other natural processes are inherent in the above mentioned systems \cite{Dissipative_transition1, Dissipative_transition2}, which give rise to various dissipative effects for sufficiently strong dissipation \cite{Dissipative_transition3,Dissipative_transition4,Dissipative_transition5,Dissipative_transition6,Dissipative_transition7}.
In a single cavity, the atom-photon interacting systems within a specific regime can be effectively described by the Jaynes-Cummings \cite{JC} or Tavis-Cummings model \cite{TC}, depending on the number of atoms in it. Moreover, coupling the cavities in an array opens up the possibility to explore many body physics with light matter interacting systems \cite{Angelakis,JCH_Plenio,Greentree,Plenio_Review1,Hartmann,Carusotto_Review,Hopping_Houck,Superfulidity_light,Vortex_Dominici,Vortex_Carusotto,Plenio_effective_spin_system,Polariton_Yamamoto,Fazio_glassy_phase,Hall_Sugato,Girvin_time_reversal}, similar to the Hubbard model. 
A variety of these models can exhibit quantum phase transitions, which have been explored theoretically \cite{Blatter,Le_Hur,Fazio_glassy_phase,M_Knap,Sibastian_1,Sugato_bose,Yamamoto_Glass}. The simplest configuration of such a many body system is the dimer of two coupled cavities forming a Jaynes-Cummings Josephson junction (JCJJ), which has been realized in circuit QED setup \cite{JC_dimer_expt}. {This system} can serve as a test bed to study various non-equilibrium phenomena \cite{JC_dimer_expt,JCD_Houck,JCD_Manus_K}. 

In the present work, we investigate the non-equilibrium dynamics and the various associated quantum phenomena in an atom-photon interacting system described by JCJJ in the presence of Kerr nonlinearity \cite{Dan_Walls,Kerr_1,Kerr_2,Kerr_3,Kerr_4,Kerr_5,Kerr_6,Kerr_7,Kerr_8}.
An insight into the overall dynamical behavior can be gained from the semiclassical analysis, which is also useful for identifying a variety of photonic Josephson oscillations in the JCJJ and transitions between them.
Interestingly, this system exhibits a self-trapping phenomenon, for which photons are dynamically localized in one of the cavities \cite{JC_dimer_expt,JCD_Houck,JCD_Sadri,JCD_Manus_K}.
Apart from this, other self trapped states also appear as a consequence of Kerr nonlinearity, which we analyze in detail, focusing on their dynamical origin and regime of stability.
On the other hand, in quantum dynamics, atoms and photons become entangled, which gives rise to interesting quantum effects, leading to the deviation from classical behavior.
Additionally, the photon field can lose its coherence as a result of phase fluctuation during the time evolution. 
%
%
It is a pertinent issue to study the entanglement dynamics and change in the state of photons due to the combined effect of interaction and entanglement for different dynamical states, as well as for a rapid quench to a dynamically unstable regime.
We also demonstrate how the self-trapping phenomena can be employed to control the photon mediated correlation between the atomic qubits, which are otherwise non interacting.
Such dynamical manipulation of entanglement between the qubits in the Jaynes-Cummings dimer model can have potential applications in quantum information processing.

The paper is organized as follows. In Sec.\ref{2}, we describe the JCJJ model and analyze it semiclassically in Sec.\ref{3} to obtain different branches of Josephson dynamics, their stability as well as transitions between them. Quantum dynamics and its comparison with semiclassical steady states are presented in Sec.\ref{4}. Sec.\ref{5} contains a detailed discussion on the quantum nature of the photon field, particularly phase diffusion and revival phenomena. In this section, we also investigate the entanglement properties of the spin $1/2$ atomic qubits corresponding to the different steady states, as well as the signature of phase space mixing of photons in the quench dynamics. Finally, we summarize the results and conclude in Sec.\ref{Conclusion}. 
         
\section{THE MODEL} \label{2}
\label{Classical}
The Jaynes-Cummings Josephson junction formed by coupling two cavities \cite{JC_dimer_expt}, can be described by the Hamiltonian, 
\begin{equation}
	\hat{\mathcal{H}}=\sum_{i}\left[\hat{\mathcal{H}}_{\rm JC}^{(i)}+\frac{U}{2}\hat{n}_i(\hat{n}_i-1)\right]- J\big(\hat{a}^{\dagger}_L\hat{a}_R+h.c.\big)-\mu\hat{M}
	\label{Hamiltonian}
\end{equation}         
where, the site index $i=L,R$ indicates the left and right cavity, which are coupled by the Josephson coupling $J$. Each cavity can be modeled by the Jaynes-Cummings Hamiltonian,
\begin{eqnarray}
\hat{\mathcal{H}}_{\rm JC}^{(i)} = \omega\hat{n}_{i}+\omega_0\hat{\sigma}^{+}_i\hat{\sigma}_i^{-}+g\big(\hat{a}_i\hat{\sigma}_{i}^++\hat{a}^{\dagger}_i\hat{\sigma}_{i}^-\big),
\end{eqnarray}
describing the interaction between an atom and single mode cavity field with frequency $\omega$, represented by the annihilation (creation) operators $\hat{a}_i$($\hat{a}^{\dagger}_i$). The two level atom with energy gap $\omega_0$ at each cavity is described by the Pauli spin operators $\vec{\hat{\sigma}}_i$. The last term of $\hat{\mathcal{H}}_{\rm JC}^{(i)}$ describes the atom-photon interaction with strength $g$. In addition, we consider the effect of Kerr nonlinearity \cite{Dan_Walls,Kerr_1,Kerr_2,Kerr_3,Kerr_4,Kerr_5,Kerr_6,Kerr_7,Kerr_8} in each cavity, represented by the second term in Eq.\eqref{Hamiltonian}, giving rise to the repulsive interaction of the photon field with strength $U$.
The JCJJ described by the Hamiltonian in Eq.\eqref{Hamiltonian} preserves the U(1) symmetry similar to the Jaynes-Cummings model \cite{Dan_Walls}, leading to the conserved total excitation number,
\begin{eqnarray}
\hat{M}=\sum_{i=L,R}(\hat{n}_i+\hat{\sigma}^{+}_i\hat{\sigma}_i^{-}).
\end{eqnarray} 
In the grand canonical ensemble, the $\mu$ in Eq.\eqref{Hamiltonian} represents the chemical potential corresponding to the number of excitations. 
Such a JCJJ has been realized in circuit QED setup \cite{JC_dimer_expt}, where the strength of interactions and the photon hopping amplitude can be tuned. 
Moving forward, we will discuss the different Josephson oscillations of the JCJJ, described by the Hamiltonian in Eq.\eqref{Hamiltonian} within the semiclassical method and compare them with the quantum mechanical dynamics.
Throughout the paper, we use units such that $\hbar,k_B=1$ and have scaled the energy (time) by $J$ $(1/J)$. 

\section{SEMICLASSICAL ANALYSIS}\label{3}
In this section, we study the dynamics of the JCJJ governed by the Hamiltonian given in Eq.\eqref{Hamiltonian}, using the time dependent variational method \cite{Dirac}.
The photons and two level atoms in the cavities can be described semiclassically by their respective coherent states \cite{Coherent_state}, using which we construct the following time dependent variational wavefunction,
\begin{eqnarray}
\ket{\psi_c(t)} = \prod_{i=L,R}\ket{\alpha_i(t)}\otimes \ket{\theta_i(t),\phi_i(t)}.
\end{eqnarray}
The coherent state of the cavity mode is given by, 
\begin{eqnarray}
\ket{\alpha_i} = \exp(\alpha_i \hat{a}_i^{\dagger}-\alpha_i^*\hat{a}_i)\ket{0}
\label{photon_coherent-state}
\end{eqnarray}
where $\alpha_i$ is the eigenvalue of $\hat{a}_i$, representing the photon field classically. The wavefunction for the two level atoms can be expressed as follows, 
\begin{eqnarray}
\ket{\theta_i,\phi_i} = \cos(\theta_i/2)\ket{\uparrow}+\sin(\theta_i/2)e^{i\phi_i}\ket{\downarrow},
\label{spin_coherent-state}
\end{eqnarray}
where $\ket{\downarrow}$($\ket{\uparrow}$) represents the ground (excited) state and the canonically conjugate variables $\phi_i,z_i = \cos\theta_i$ describe the orientation of such a spin 1/2 system on the Bloch sphere, for which $\langle \vec{\hat{S}}_i\rangle = S(\sin\theta_i\cos\phi_i,\sin\theta_i\sin\phi_i,\cos\theta_i) $ with $S=1/2$.
The coherent state representation of the photon field is appropriate for a large number of photons in each cavity, giving rise to the substantial number of conserved total excitations, that can be written semiclassically as $M = \sum_i|\alpha_i|^2+(1+z_i)/2$. 
It is evident from the conservation equation that the amplitude of the classical field $\alpha_i$ scales with $\sqrt{M}$. Therefore, for a large number of conserved excitations, we define $\alpha_i/\sqrt{M} = \sqrt{n_i}\exp(\iota \psi_i)=(x_i+\iota p_i)/\sqrt{2}$, where, $n_i\in [0,1]$ is the scaled photon number, $\psi_i$ represents its phase, and $x_i,p_i$ are the corresponding conjugate variables.
In terms of the dynamical variables $\mathbf{x}=\{n_i,\psi_i,z_i,\phi_i\}$, the Lagrangian scaled by the total excitation number $M$ can be written as, 
\begin{eqnarray}
	\mathcal{L}&=& \frac{1}{M}\bra{\psi_c}i\frac{\partial}{\partial t}-\hat{\mathcal{H}}\ket{\psi_c}\nonumber\\
	&=&\sum_{i=L,R}\left[-\dot{\psi}_in_i+\frac{\eta}{2}\dot{\phi}_iz_i-(\omega-\mu) n_i-\frac{\eta}{2}(\omega_0-\mu)z_i\right.\nonumber\\
	&&-\frac{\tilde{U}}{2}n_i^2
	-\left.\tilde{g}\sqrt{n_i}\sqrt{1-z_i^2}\cos(\phi_i+\psi_i)\right]\nonumber\\
	&&+2\sqrt{n_Ln_R}\cos(\psi_L-\psi_R),
\end{eqnarray}
where $\eta=2S/M$ and the interaction strengths are scaled as $\tilde{g} = g/\sqrt{M}$, $\tilde{U}=UM$. Note that, in general $\eta=2S/M$ for a large spin system with magnitude $S$, which is considered to be small in the present case of the Jaynes-Cummings model with $S=1/2$ and $M\gg1$.
From the Euler-Lagrange equation $\frac{d}{dt}\left(\frac{\partial\mathcal{L}}{\partial\dot{\textbf{x}}}\right)-\frac{\partial\mathcal{L}}{\partial\textbf{x}}=0$ of the dynamical variables $\textbf{x} = \{n_i,\psi_i,z_i,\phi_i\}$, we obtain the following equations of motion (EOM),
\begin{subequations}
	\begin{eqnarray}
	\dot{n_i}&=&-\tilde{g}\sqrt{n_i}\sqrt{1-z_i^2}\sin{(\phi_i+\psi_i)}\notag\\
	&&+2\sqrt{n_in_{\overline{i}}}\sin{(\psi_i-\psi_{\overline{i}})} \\
	\dot{\psi_i}&=& -(\omega-\mu)-\frac{\tilde{g}}{2\sqrt{n_i}}\sqrt{1-z_i^2}\cos{(\phi_i+\psi_i)} \notag \\
	 &&+\sqrt{\frac{n_{\overline{i}}}{n_i}}\cos{(\psi_i-\psi_{\overline{i}})}-\tilde{U}n_i  \\
	\eta\dot{\phi_i}&=& \eta(\omega_0-\mu)-\frac{2\tilde{g}z_i}{\sqrt{1-z_i^2}}\sqrt{n_i}\cos{(\phi_i+\psi_i)}\\
	\eta\dot{z_i}&=&2\tilde{g}\sqrt{n_i}\sqrt{1-z_i^2}\sin{(\phi_i+\psi_i)} 
	\end{eqnarray}
	\label{EOM}
\end{subequations}
where $\bar{i}\neq i$. Conservation of the total excitation number yields the constraint,
\begin{eqnarray}
n_L+n_R+\frac{\eta}{2}\big(z_L+z_R+2\big)=1.
\label{Conservation_equation} 
\end{eqnarray}
We solve Eq.\eqref{EOM} within the grand canonical ensemble, where $\mu$ is fixed by the Eq.\eqref{Conservation_equation}. 
However, in the limit $\tilde{g}\rightarrow 0$, both the photon number and atomic inversion become conserved individually and therefore, our formalism can not be continued to this limit. Hence, we exclude the regime of small $\tilde{g}$ from our discussion.
First, we investigate the steady states corresponding to the fixed point (FP) $\mathbf{x}^*= (n_i^*,\psi_i^*,z_i^*,\phi_i^*)$ of the EOM given in Eq.\eqref{EOM}, for which $\dot{\mathbf{x}}=0$.
Next, we perform the linear stability analysis around the steady states, describing the evolution of small initial fluctuation $\delta\mathbf{x}(0)$ in the form $\delta\mathbf{x}(t) = \delta\mathbf{x}(0)e^{i \tilde{\omega} t}$ and determine the frequency $\tilde{\omega}$.
The stability of  FPs is ensured if the  $\rm Im(\tilde{\omega}) = 0$ and the $\tilde{\omega}$ yields the small amplitude oscillation frequency around the corresponding steady states.
In the JCJJ, the stable steady states describe the different types of photonic Josephson oscillations with the frequency that can be obtained from the linear stability analysis mentioned above. Next, we find the different possible steady states from Eq.\eqref{EOM} and analyze their stability.

\subsection{STEADY STATE ANALYSIS}\label{3a}
\begin{figure}[h]
	\centering
	\includegraphics[width=0.95\columnwidth]{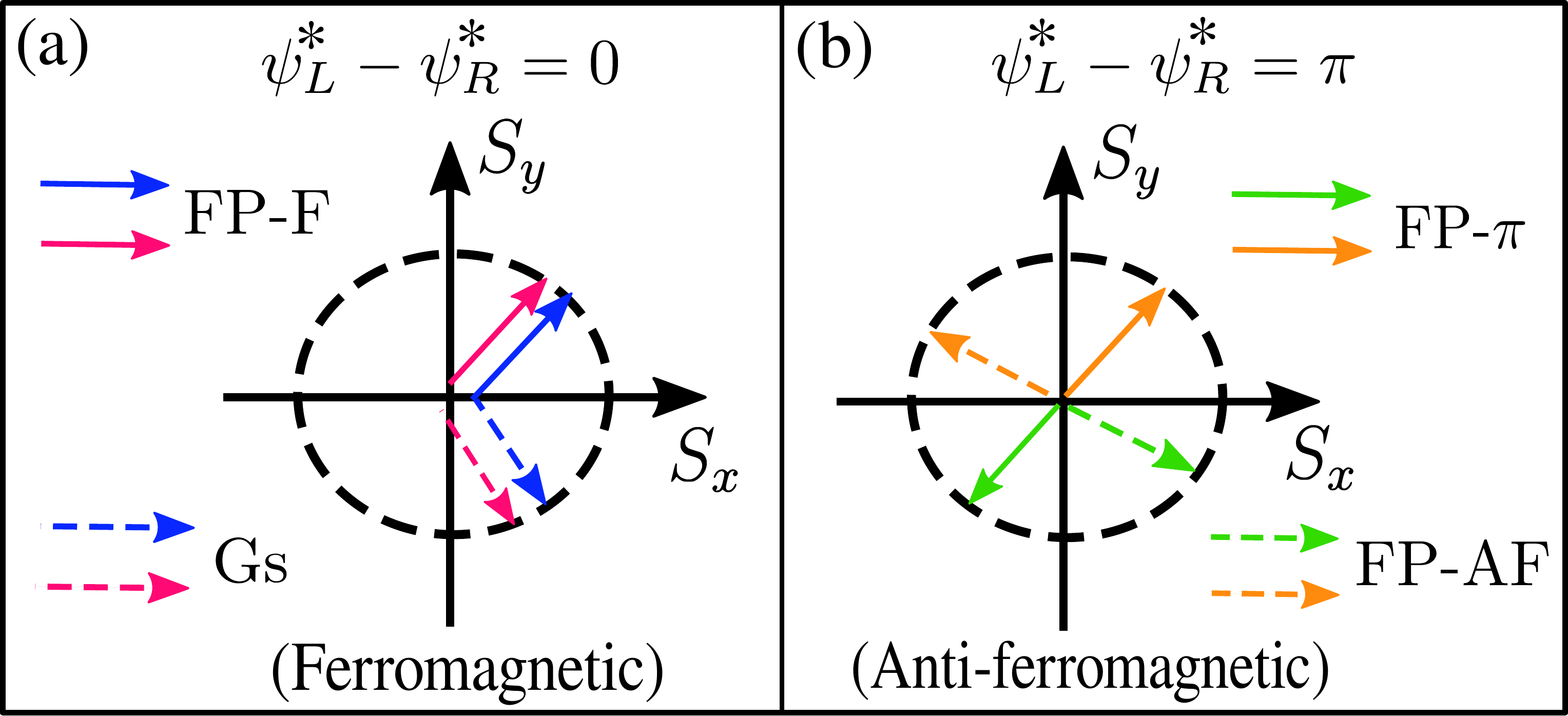}
	\caption{{\it Schematic diagram illustrating the spin orientations of the various steady states in the $S_x$-$S_y$ plane:}  (a) depicts the steady states corresponding to the ferromagnetic class, and (b) represents the steady states corresponding to the antiferromagnetic class. }	
	\label{fig1}
\end{figure}

In this subsection, we systematically investigate various steady states obtained from the EOM in Eq.\eqref{EOM} and analyze their stability as outlined above. As evident from Eq.\eqref{EOM}(a,d), the steady states satisfy the conditions,
\begin{subequations}
	\begin{eqnarray}
		\sin(\phi^*_i+\psi^*_i)&=&0\\
		\sin(\psi^*_L-\psi^*_R)&=&0,
	\end{eqnarray}
\end{subequations}
which correspond to the phase relations $\phi^*_i+\psi^*_i=0,\pi$ and $\psi^*_L-\psi^*_R=0,\pi$, that is used to classify the steady states.
The relative phase of bosons $\psi^*_L-\psi^*_R=0(\pi)$
equivalently describes the (anti)ferromagnetic spin configuration of the cavities in the $S_x$-$S_y$ plane, corresponding to $\phi_L^*-\phi_R^*=0(\pi)$. We categorize the steady states in these two classes, which are represented schematically in Fig.\ref{fig1}(a,b).
Note that the transformations $\phi^*_i\rightarrow\phi^*_i+\delta$ and $\psi^*_i\rightarrow\psi^*_i-\delta$ leave the steady state equations Eq.\eqref{EOM} invariant as a consequence of the U(1) symmetry. This results in a continuous set of FPs lying on circles in the $x_i$-$p_i$ and $S_{ix}$-$S_{iy}$ planes with corresponding radius $\sqrt{2n^*_i}$ and $\sqrt{1-z_i^{*2}}/2$, respectively (see Fig.\ref{fig1}(a,b)). 
For a particular class of spin configuration and a given value of $\eta$, the steady states can be obtained in terms of $\{n_i^*,z_i^*\}$, by solving Eq.\eqref{EOM}(b,c), subjected to the constraint in Eq.\eqref{Conservation_equation}, which conserves the total excitation. The steady states thus obtained, can be categorized in terms of the relative photon population $f =  n^*_R/n^*_L$, which we denote as $symmetric$ ($f=1$) and $self\,\,trapped$ ($f\ne 1$), corresponding to equal and unequal photon population in the cavities. Note that, once the photon population $n_i^*$ is obtained, it also determines the atomic inversion $z_i^*$,
\begin{eqnarray}
z^*_i = \frac{\xi_2\eta(\omega_0-\mu)}{\sqrt{\eta^2(\omega_0-\mu)^2+4\tilde{g}^2n^*_i}}.
\label{z_equation}	
\end{eqnarray}
It is important to mention that the Hamiltonian given in Eq.\eqref{Hamiltonian} remains invariant under the exchange of the degrees of freedom between the two cavities ($\hat{a}_L\leftrightarrow \hat{a}_R$ and $\hat{\vec{\sigma}}_L\leftrightarrow \hat{\vec{\sigma}}_R$), which indicates the discrete left-right symmetry between the cavities. The spontaneous breaking of this symmetry can give rise to the self trapped state.
Next, we analyze the steady state equations graphically, which provides a physical picture and qualitative behavior of the steady states as well as the transitions \cite{Dyn_transition1,Dyn_transition2} between them. For small values of $\eta$, from Eq.\eqref{Conservation_equation}, the total photon number can be approximately written as $n_L^*+n_R^*=1-\eta$, which yields,
\begin{eqnarray}
n_L^*=\frac{(1-\eta)}{1+f},\quad n_R^*=\frac{f(1-\eta)}{1+f}.
\end{eqnarray}
Using these relations, the steady state equations Eq.\eqref{EOM}(b,c) can be reduced to a single effective equation in terms of the relative photon population $f$,
\begin{align}
\mathcal{Y}(f) =\,\,&\xi_1(f-1)-\tilde{U}(1-\eta)\left(\frac{1-f}{f+1}\right)\sqrt{f}\nonumber\\
&-\,\xi_2\tilde{g}^2\sqrt{f}\left(\frac{1}{\sqrt{\mathcal{F}_L(f)}}-\frac{1}{\sqrt{\mathcal{F}_R(f)}}\right)=0,
\label{self-consistent_equation}
\end{align} 
where, $\mathcal{F}_i(f) = \eta^2(\omega_0-\mu)^2+4\tilde{g}^2n_i^*(f)$ and the discrete variable $\xi_1 = \cos(\psi_L^*-\psi_R^*) =\pm 1$ describes the spin orientation of two qubits while $\xi_2 = \cos(\phi^*_i+\psi_i^*) = \pm 1$.
For small $\eta$, the chemical potential $\mu$ is given by,
\begin{align}
\mu = \omega-\frac{\xi_1}{2}\left(\sqrt{f}+\frac{1}{\sqrt{f}}\right)+\frac{\tilde{U}}{2}+\frac{\xi_2\tilde{g}}{4}\sqrt{1+f}\left(1+\frac{1}{\sqrt{f}}\right).
\label{mu}
\end{align}
Note that, as a consequence of the left-right symmetry between the cavities,  Eq.\eqref{self-consistent_equation} remains invariant under the transformation $f\rightarrow1/f$, hence we only consider the steady state solutions for $f\in [0,1]$.
The roots of Eq.\eqref{self-consistent_equation} yield the possible steady states for a given combination of $\xi_1,\xi_2$, which we discuss below.
\subsubsection{Ferromagnetic class $(\boldsymbol{\psi_L^*-\psi_R^*=0})$ }
\begin{figure}[h]
	\centering
	\includegraphics[width=\columnwidth]{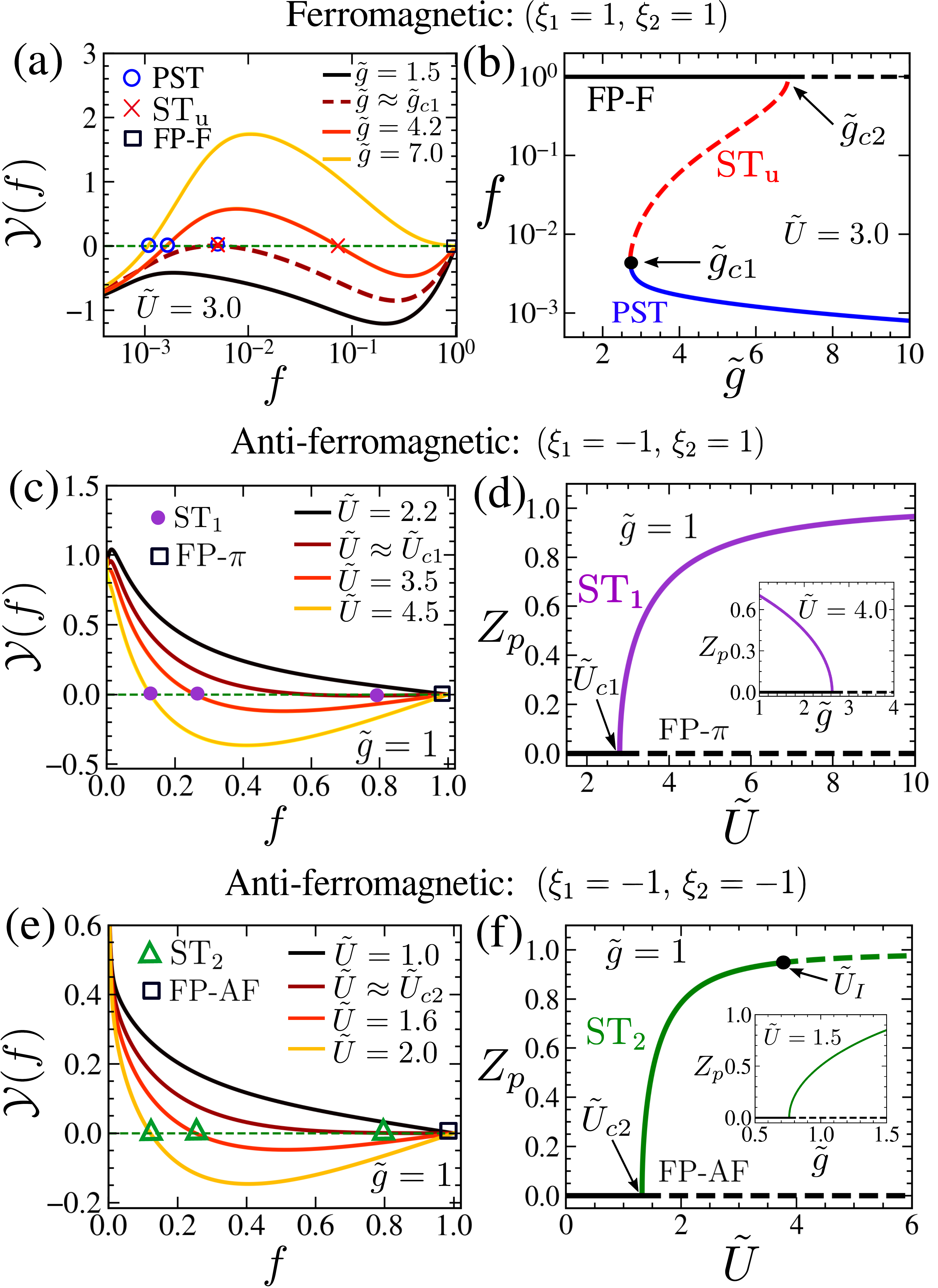}
	\caption{{\it Graphical analysis of the function $\mathcal{Y}(f)$ to identify different steady states and their bifurcations:} 
	(a) Identification of the steady states from the roots of the function $\mathcal{Y}(f)$ and (b) the steady state solutions in terms of $f$ (see the text for details) are shown as a function of $\tilde{g}$ for $\tilde{U}=3$, corresponding to the ferromagnetic class ($\xi_1=1,\xi_2=1$). (c,e) Graphical representation of $\mathcal{Y}(f)$ and (d,f) bifurcation diagram of the FPs in terms of photon population imbalance $Z_p$ for anti-ferromagnetic class ($\xi_1=-1$) corresponding to $\xi_2=\pm 1$, respectively (see the main text), at $\tilde{g}=1$ for different Kerr nonlinearity $\tilde{U}$.  The different markers in (a,c,e) indicate the various steady states corresponding to the respective classes. Solid (dashed) lines in (b,d,f) represent stable (unstable) steady states.
	Inset of (d,f) shows the behavior of $Z_p$ as a function of $\tilde{g}$ for $\tilde{U}=4,1.5$ respectively.
	Here and in rest of the figures, we consider excitation number $M=30$, equivalently, $\eta=1/M$, unless otherwise mentioned. The energy $E$ and interaction strengths $\tilde{g},\tilde{U}$ are measured in the units of hopping amplitude $J$. We set $\hbar,k_B = 1$ and $\omega=\omega_0=2$ for all figures.}
	\label{fig2}
\end{figure}

For the ferromagnetic orientation of the qubits, $\xi_1=+1$ while the other variable can take two values $\xi_2 = \pm 1$. When $\xi_2=-1$, the equation $\mathcal{Y}(f)$ has only one root for $f=1$, describing a symmetric steady state corresponding to the ground state configuration (Gs). On the other hand, $\xi_2=1$ is a more interesting scenario since it gives rise to various non trivial steady states, as shown in Fig.\ref{fig2}(a,b). Similar to the previous case, $f=1$ is always a solution of equation $\mathcal{Y}(f)$ describing a symmetric state with higher energy density (scaled by the total number of excitation), which is denoted by FP-F. 
Interestingly, two new solutions appear above a critical coupling  strength,
\begin{eqnarray}
	\tilde{g}_{c1}(\tilde{U})= 2+3\left(\frac{1+\tilde{U}}{2}\right)^{4/3}\eta^{2/3}-\frac{\eta}{2},
	\label{gc1}
\end{eqnarray}
giving rise to two self trapped states, one of which is unstable, as seen from Fig.\ref{fig2}(b). The FP with vanishingly small relative photon population $f\approx 0$ corresponds to a stable {\it perfect self trapped} (PST) state \cite{JC_dimer_expt,JCD_Houck}, describing a situation where almost all the photons are localized in one of the cavities. As illustrated in Fig.\ref{fig2}(a,b), such self trapped states arise as a result of a saddle-node bifurcation occurring at $\tilde{g}_{c1}(\tilde{U})$, for which non vanishing small parameter $\eta$ plays a crucial role. 
The unstable self trapped state ST$_{\rm u}$ (with larger value of $f$) undergoes a subcritical pitchfork bifurcation with FP-F at the critical point $\tilde{g}_{c2}(\tilde{U})$, which can be approximately written as,
\begin{eqnarray}
	\tilde{g}_{c2}(\tilde{U}) = \left(\sqrt{8}+\sqrt{2}\tilde{U}\right)-\left(\frac{3}{\sqrt{2}}\tilde{U}+\sqrt{2}\right)\eta,	
	\label{gc2}
\end{eqnarray}
after which the symmetric state FP-F becomes unstable, as depicted in Fig.\ref{fig2}(b). 
Now we focus on the steady states corresponding to the anti-ferromagnetic spin configuration.
\subsubsection{Anti-ferromagnetic class $(\boldsymbol{\psi_L^*-\psi_R^*=\pi})$ }
For the anti-ferromagnetic class with $\xi_1=-1$, the steady states at higher energies compared to the ground state and the transitions between them are very intriguing where the Kerr nonlinearity $\tilde{U}$ plays a crucial role. The other variable can take two values $\xi_2=\pm 1$, and we discuss the corresponding steady states one by one.

$\boldsymbol{\xi_2=1:}$ In this case, a symmetric steady state denoted by FP-$\pi$ exists, which undergoes a pitchfork bifurcation at a critical Kerr nonlinearity, as evident from Fig.\ref{fig2}(c,d). After the bifurcation, FP-$\pi$ becomes unstable, giving rise to a stable self trapped state ST$_1$. This bifurcation is indicated by the solid black line in Fig.3(b). This phenomenon also occurs in the Bose-Josephson junction, in absence of coupling to the spin ($\tilde{g}=0$) \cite{Shenoy1,Shenoy2,vardi1,vardi2}, which has been detected experimentally \cite{Oberthalar,BJJ_self_trapped}. However, in the present case, the critical Kerr nonlinearity also depends on the coupling strength $\tilde{g}$, which is given by,
\begin{eqnarray}
	\tilde{U}_{c1}(\tilde{g}) = 2+\frac{\tilde{g}}{\sqrt{2}}+\left(2+\frac{3\tilde{g}}{2\sqrt{2}}\right)\eta,	
	\label{uc1}
\end{eqnarray}
for small $\eta$. Unlike the perfect self-trapping, the relative photon population imbalance between the two cavities, 
\begin{eqnarray}
Z_p = \frac{n_L-n_R}{n_L+n_R}
\label{Zp}
\end{eqnarray}
of the self trapped state ST$_1$ increases continuously after the bifurcation and approaches to unity with increasing Kerr nonlinearity $\tilde{U}$, which is shown in Fig.\ref{fig2}(d). In contrast, the relative population imbalance $Z_p$ decreases with increasing atom-photon coupling strength $\tilde{g}$ (see the inset of Fig.\ref{fig2}(d)), which serves as a characteristic feature of this ST$_1$ state for its identification.
It is evident from Eq.\eqref{z_equation}, for the self trapped state, the photon population imbalance $Z_p$ leads to the atomic population imbalance,
\begin{eqnarray}
Z_a = \frac{|z_R-z_L|}{2},
\label{Za}
\end{eqnarray}
exhibiting similar behavior with coupling strength.

\begin{figure}
	\centering
	\includegraphics[width=\columnwidth]{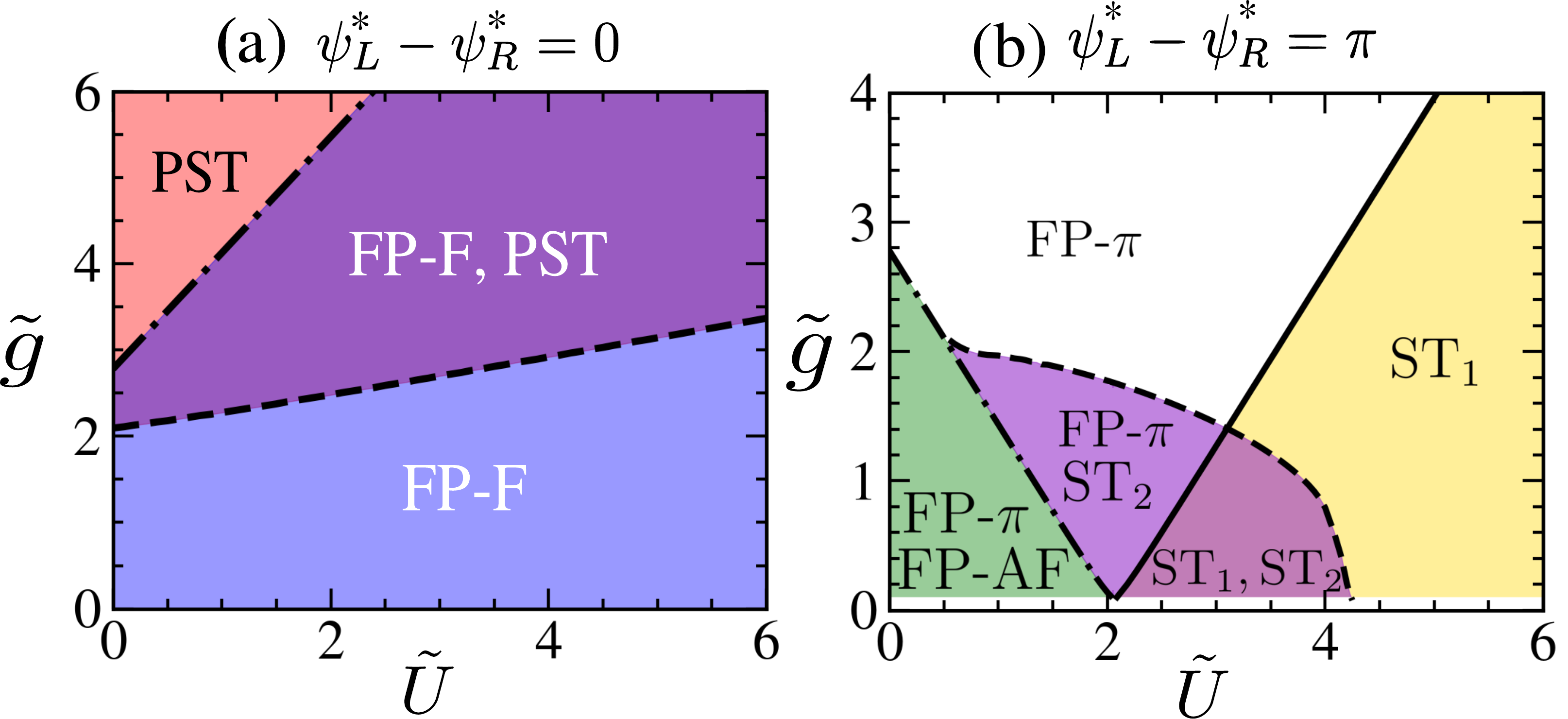}
	\caption{{\it Steady state phase diagram as a function of interaction strengths $\tilde{U}$ and $\tilde{g}$:} (a) Ferromagnetic and (b) anti-ferromagnetic class. Here, all the phase boundaries are obtained numerically. The boundary of PST (dashed line) and instability line of FP-F (dashed-dotted line) in (a) are described approximately by $\tilde{g}_{c1}$ and $\tilde{g}_{c2}$, respectively, see Eq.(\ref{gc1},\ref{gc2}) of the main text. The bifurcation lines between FP-AF, ST$_2$ (dashed-dotted line) and FP-$\pi$, ST$_1$ (solid line) in (b) can be written approximately using $\tilde{U}_{c1}$ and $\tilde{U}_{c2}$, as given by Eq.(\ref{uc1},\ref{uc2}) in the main text. The ST$_2$ state becomes unstable outside the dashed line ($\tilde{U}_{\rm I}(\tilde{g})$) without giving rise to any new steady state. Note that, the small $\tilde{g}$ regime is kept blank, since our formalism can not be continued to $\tilde{g}=0$ (see the text for details).
	}
	\label{fig3}
\end{figure}

$\boldsymbol{\xi_2=-1:}$ A similar type of phenomenon can also be observed for $\xi_2=-1$.
In this case, another symmetric state FP-AF exists, which is energetically different from FP-$\pi$ but corresponds to the same anti-ferromagnetic spin orientation. 
The symmetric state FP-AF undergoes a pitchfork bifurcation at a critical strength of Kerr interaction,
\begin{eqnarray}
	\tilde{U}_{c2}(\tilde{g}) = 2-\frac{\tilde{g}}{\sqrt{2}}+\left(2-\frac{3\tilde{g}}{2\sqrt{2}}\right)\eta,	
	\label{uc2}
\end{eqnarray}
which occurs only for $\tilde{g}\lesssim 2$. Above this critical coupling, the FP-AF state becomes unstable, giving rise to a new self trapped state denoted by ST$_2$ (see Fig.\ref{fig2}(e,f)). Unlike ST$_1$, this self trapped state loses its stability above a critical Kerr nonlinearity $\tilde{U}_{\rm I}(\tilde{g})$ (denoted by the black dashed line in Fig.\ref{fig3}(b)) without forming a new steady state, as shown in the bifurcation diagram given in Fig.\ref{fig2}(f) for a fixed coupling $\tilde{g}$.	
Thus ST$_2$ can exist as a stable state only in the range $\tilde{U}_{c2}(\tilde{g})\le\tilde{U}<\tilde{U}_{\rm I}(\tilde{g})$ and for $\tilde{g}\lesssim 2$.
The relative photon population imbalance $Z_p$ for ST$_2$ state increases and approaches unity as both the interaction strengths $\tilde{U},\tilde{g}$ increases (as depicted in Fig.\ref{fig2}(f)). This behavior is strikingly different from that of the ST$_1$, where $Z_p$ diminishes with $\tilde{g}$. Such qualitatively different features can be employed to distinguish between the two self trapped states ST$_1$ and ST$_2$ during quantum dynamics, which we will discuss in the next section. Note that, in addition to ST$_1$ and ST$_2$, other self trapped states can also appear, exhibiting complicated scenarios, which we prefer to leave out from the present discussion as they are less relevant due to their existence within a small range of parameters. Moreover, the signatures of these states have not been found in quantum dynamics.
 
In the limit $\tilde{g}\rightarrow 0$, the steady states corresponding to $\xi_2=\pm 1$ become almost identical (see Eq.\eqref{self-consistent_equation}), with a small difference of the order of $\eta$ in the physical quantities. In this regime, both the self trapped states ST$_1$ and ST$_2$ become practically identical. However, we exclude the small $\tilde{g}$ regime from our discussion, as the formalism can not be extrapolated to $\tilde{g} =0$, for which both the atomic excitation and photon number are conserved separately.

The plethora of steady states obtained from the above analysis are summarized in the phase diagrams, depicted in Fig.\ref{fig3}(a,b), separately for ferromagnetic and anti-ferromagnetic classes, indicating their region of stability. Here, the phase diagrams are obtained by solving the steady state equations exactly for a fixed value of $\eta$ (equivalently, a fixed number of excitations $M$). The numerically obtained phase boundaries of the steady states PST, FP-F, (shown in Fig.\ref{fig3}(a)) and the transition lines between FP-$\pi$ to ST$_1$ as well as FP-AF to ST$_2$, (depicted in Fig.\ref{fig3}(b)) are in good agreement with the analytical results given in Eq.(\ref{gc1},\ref{gc2},\ref{uc1},\ref{uc2}) for small values of $\eta$. 
The appropriate parameter regimes can be identified from the phase diagrams for observation of different dynamical behavior and transitions. 

\subsection{CLASSICAL DYNAMICS}
\begin{figure}
	\centering
	\includegraphics[width=\columnwidth]{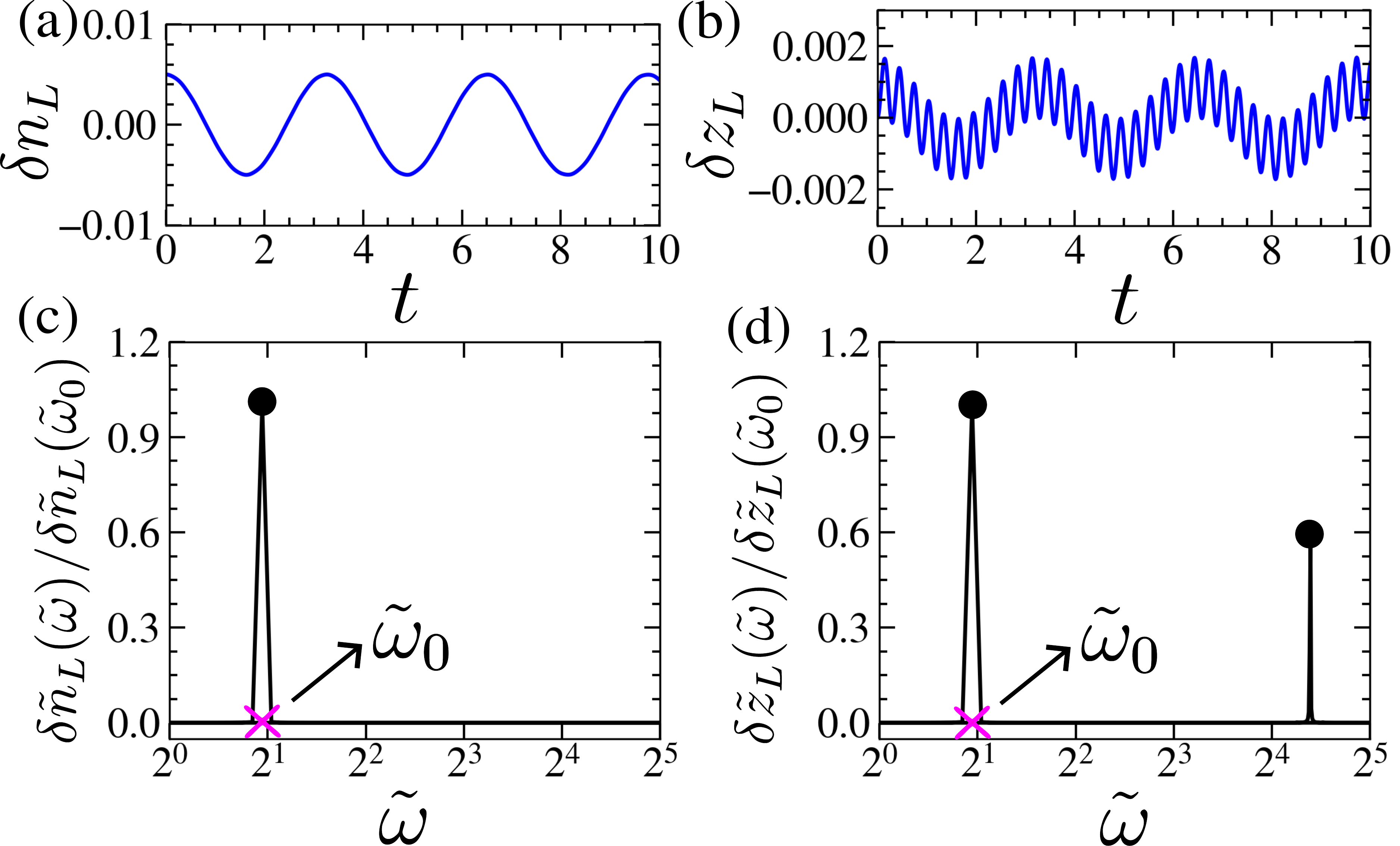}
	\caption{{\it Small amplitude oscillations of the different physical quantities and their characteristic frequencies corresponding to FP-$\pi$ state:} Dynamics of (a) photon number $\delta n_L(t)=n_L(t)-n_L^*$, (b) atomic inversion $\delta z_L(t)=z_L(t)-z_L^*$, subtracted from their steady state values. The respective Fourier spectra, $\delta \tilde{n}_L(\tilde{\omega})/\delta \tilde{n}_L(\tilde{\omega}_0)$ and $\delta \tilde{z}_L(\tilde{\omega})/\delta \tilde{z}_L(\tilde{\omega}_0)$, scaled by the amplitude of the lowest frequency are shown in (c) and (d). The lowest frequency $\tilde{\omega}_0$ obtained from the linear stability analysis is present in both the quantities. Here and in the remaining figures, time $t$ and frequencies $\tilde{\omega},\tilde{\omega}_0$ are scaled by $1/J$ and
		$J$ respectively.}
	\label{fig4}
\end{figure}

To this end, we investigate the classical dynamics corresponding to the different steady states illustrated in the phase diagram of Fig.\ref{fig3}, which provides useful information about various photonic Josephson oscillations and transitions between them.  The time evolution is performed by solving the EOM given in Eq.\eqref{EOM} numerically for an appropriately chosen initial condition. 
In general, if the initial condition is chosen close to a stable fixed point, the photon number and other physical quantities oscillate around the steady state, with oscillation frequencies obtained from the linear stability analysis. We illustrate the oscillation around the symmetric state FP-$\pi$ by computing the deviation of photon number $\delta n_i(t)=n_i(t)-n_i^*$ and atomic inversion $\delta z_i(t)=z_i(t)-z_i^*$ from the corresponding steady state values, which exhibits small amplitude oscillation around zero,  shown in Fig.\ref{fig4}(a,b).
Numerically, the Fourier transform of the time evolution of photon population and atomic inversion yields the relevant frequencies present in the dynamics. As observed from 
Fig.\ref{fig4}(c,d), the lowest frequency $\tilde{\omega}_0$ obtained from the linear stability analysis of the steady state FP-$\pi$ corresponds to the highest amplitude of the Fourier transform, indicating its dominant role in both the photon and spin (atom) dynamics. However, as evident from Fig.\ref{fig4}(d), the higher frequency modes also contribute to the spin degree with small amplitude, resulting in fast dynamics (see Fig.\ref{fig4}(b)).  Such dynamics around a stable fixed point as depicted in Fig.\ref{fig4}(a,b), usually lie on an invariant KAM torus \cite{KAM} with off resonant frequencies. Consequently, a trajectory over a long duration of time densely fills the region around the ring of FPs.
\begin{figure}[t]
	\centering
	\includegraphics[width=\columnwidth]{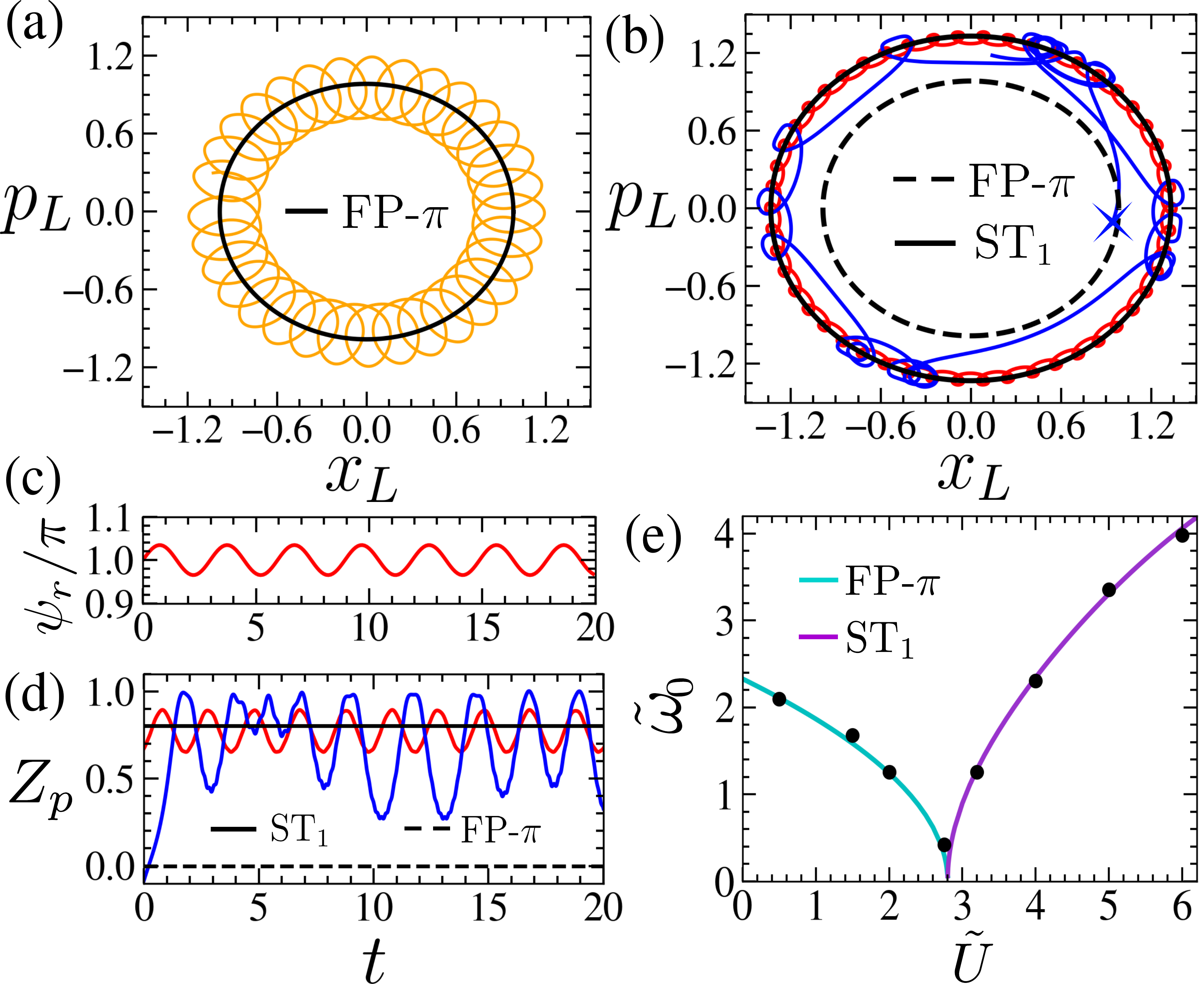}
	\caption{\textit{Classical phase portrait and variation of oscillation frequency across the transition between FP-$\pi$ and ST$_1$ states:} Dynamics of photon field in $x_L$-$p_L$ plane of left cavity corresponding to the (a) stable regime with $(\tilde{g},\tilde{U})=(1.0, 1.5)$ and (b) unstable regime with $(\tilde{g}, \tilde{U})=(1.0, 6.5)$, of FP-$\pi$ state. The red (blue) trajectory in (b) is obtained by starting the dynamics near the stable ST$_1$ (unstable FP-$\pi$) state. (c) Time evolution of the relative phase of photon $\psi_r=\psi_L-\psi_R$ for stable FP-$\pi$ state. (d) Dynamics of the relative photon population imbalance $Z_p$ in the unstable regime of FP-$\pi$. The colored lines carry the same meaning as in (b). (e) Variation of the lowest frequency $\tilde{\omega}_0$ (of small amplitude oscillation) with $\tilde{U}$ across the transition, for $\tilde{g}=1$.  The black circles and the solid lines denote the lowest frequency obtained from the Fourier transform of the dynamics, and the linear stability analysis, respectively.  	
	}
	\label{fig5}
\end{figure} 

It is fascinating to study the dynamics across the bifurcation of the steady states, particularly the emergence of the self trapped states. Here, we focus on the classical dynamics across the pitchfork bifurcation of the symmetric state FP-$\pi$ to the self trapped state ST$_1$, which occurs by tuning the Kerr nonlinearity $\tilde{U}$. Before the bifurcation, since the stable FP-$\pi$ is a symmetric state, we study the dynamics of the photon field in $x$-$p$ plane for one of the cavities, depicted in Fig.\ref{fig5}(a). 
As mentioned before, due to the U(1) symmetry, the continuous FPs lie on a circle in the $x$-$p$ plane of the photon field (black line in Fig.\ref{fig5}(a)).
Ideally, the small amplitude dynamics is expected to be confined around one of the FPs, which only occurs if we consider the initial condition for $\phi+\psi$ to be the same as the value of the FPs, without any fluctuation around it.
However, for an arbitrary initial condition around one of the FPs, the trajectory surrounds all the fixed points on the ring, as depicted in Fig.\ref{fig5}(a). As the main characteristic feature of the FP-$\pi$ mode, the relative phase of photons $\psi_r = \psi_L-\psi_R$ oscillates around the value $\pi$, which is shown in Fig.\ref{fig5}(c).
Above the critical coupling $\tilde{U}_{c1}$, FP-$\pi$ becomes unstable, and depending on the initial condition, the dynamics is attracted towards one of the stable self trapped states. It is evident from Fig.\ref{fig5}(b) that the trajectory is repelled from the FP-$\pi$ state and attracted towards the ring of FPs corresponding to the ST$_1$ state. Consequently, the photon imbalance $Z_p$ oscillates around a finite value corresponding to the steady state (see Fig.\ref{fig5}(d)).
The signature of this transition can be observed from the oscillation frequencies of FP-$\pi$ and ST$_1$ state, both of which vanish at the critical coupling strength $\tilde{U}_{c1}$, as evident from Fig.\ref{fig5}(e). A similar phenomenon also occurs for the bifurcation of FP-AF to ST$_2$ state. Such behavior is similar to the {\it mode softening} phenomenon associated with the quantum phase transition \cite{Mode_softening}. 

\begin{figure}[t]
	\centering
	\includegraphics[width=\columnwidth]{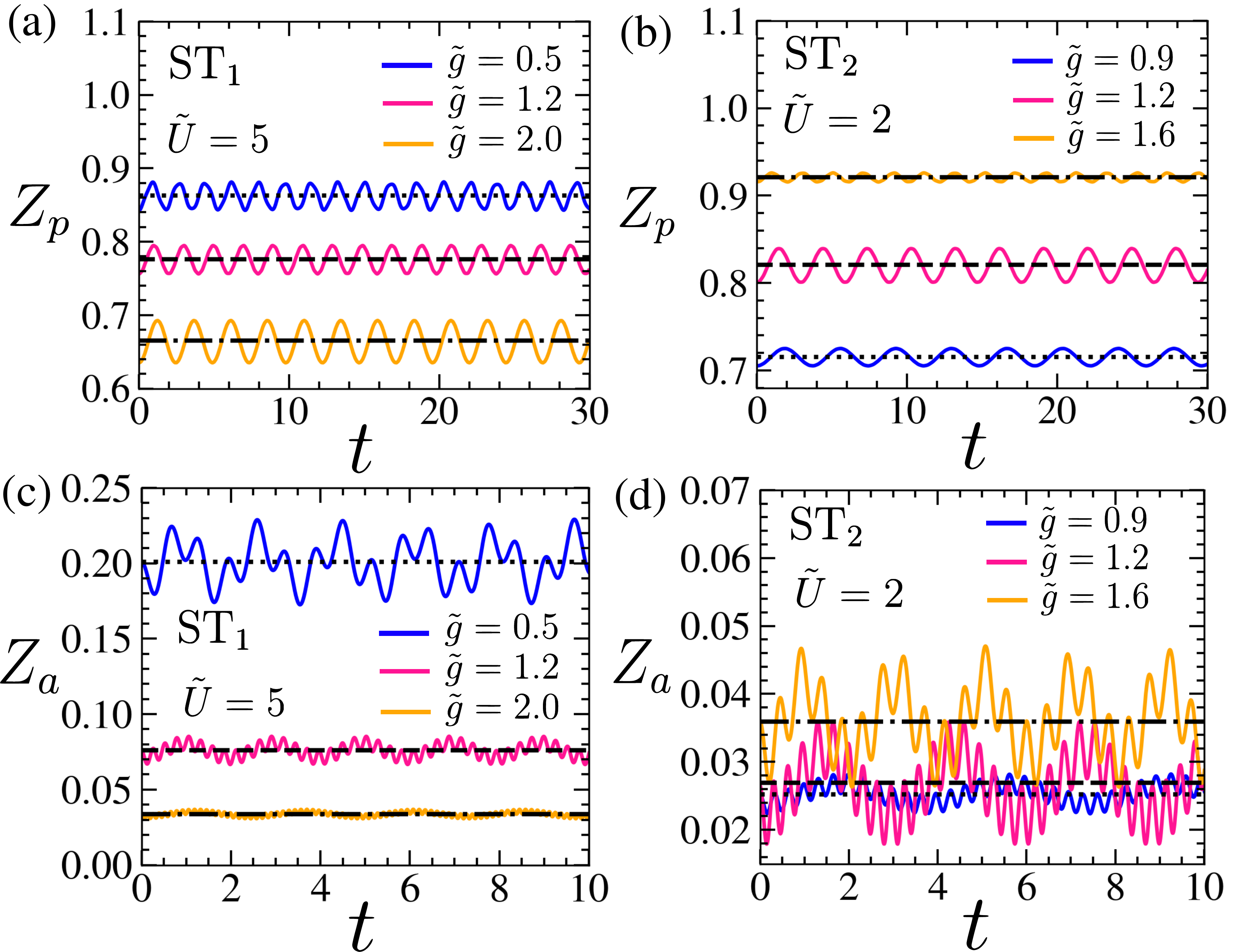}
	\caption{{\it Classical dynamics around different self trapped states:} (a,b) Relative photon population imbalance $Z_p$ and (c,d) atomic imbalance $Z_a$ for the ST$_1$ and ST$_2$ states, respectively, at different coupling strengths $\tilde{g}$. The Kerr interaction strength for the dynamics of ST$_1$ and ST$_2$ are $\tilde{U}=5$ and $\tilde{U}=2$, respectively. The horizontal lines with different line styles indicate the classical steady-state values of both $Z_p$ as well as $Z_a$. }
	\label{fig6}
\end{figure}

Next, we focus on the dynamics of the self trapped states ST$_1$ and ST$_2$. As a distinguishing feature between them, the relative photon population imbalance $Z_p$  decreases with increasing atom-photon coupling $\tilde{g}$ for ST$_1$ (see Fig.\ref{fig6}(a)) whereas it increases for ST$_2$, as depicted in Fig.\ref{fig6}(b).
%
Since the atomic inversion is directly related to the photon population in each cavity, as given in Eq.\eqref{z_equation}, the relative photon population imbalance $Z_p$ can also induce an atomic inversion imbalance, $Z_a$ for the self trapped states. The variation of $Z_a$ with $\tilde{g}$ can also distinguish between the two self trapped states ST$_1$ and ST$_2$, exhibiting opposite behavior, which is illustrated in Fig.\ref{fig6}(c,d). However, its variation is small for the ST$_2$ state as compared to that of ST$_1$.

So far, we have analyzed the classical dynamics based on a simplified description, neglecting the atom-photon correlation. Hence, it is important to investigate the signature of such dynamical states in quantum dynamics and the effect of atom-photon entanglement, which we consider in the next sections.  

\section{QUANTUM DYNAMICS}\label{4}
In this section, we study the full quantum dynamics of the JCJJ and compare them with the classical dynamics to investigate the effect of Kerr nonlinearity as well as the atom-photon correlation. We evolve the initial state $\ket{\Psi(0)}$, with a fixed number of excitations $M$, within the Schr\"{o}dinger prescription, which is performed numerically by truncating the basis up to a sufficiently large number $N_{\rm max}$. To compare with classical dynamics, we choose the initial state as the product of coherent states of photons and spins, described in Eq.(\ref{photon_coherent-state},\ref{spin_coherent-state}) respectively which represents the classical phase space point. To investigate the signature of different branches of the dynamical states, we time evolve the appropriately chosen initial state and obtain the dynamics of different physical quantities such as the population of photons and the atom in different cavities as well their imbalance, characterizing those states. The parameters are also chosen from the stability region of the corresponding states from the phase diagram, given in Fig.\ref{fig3}.

\begin{figure}[t]
	\centering
	\includegraphics[width=\columnwidth]{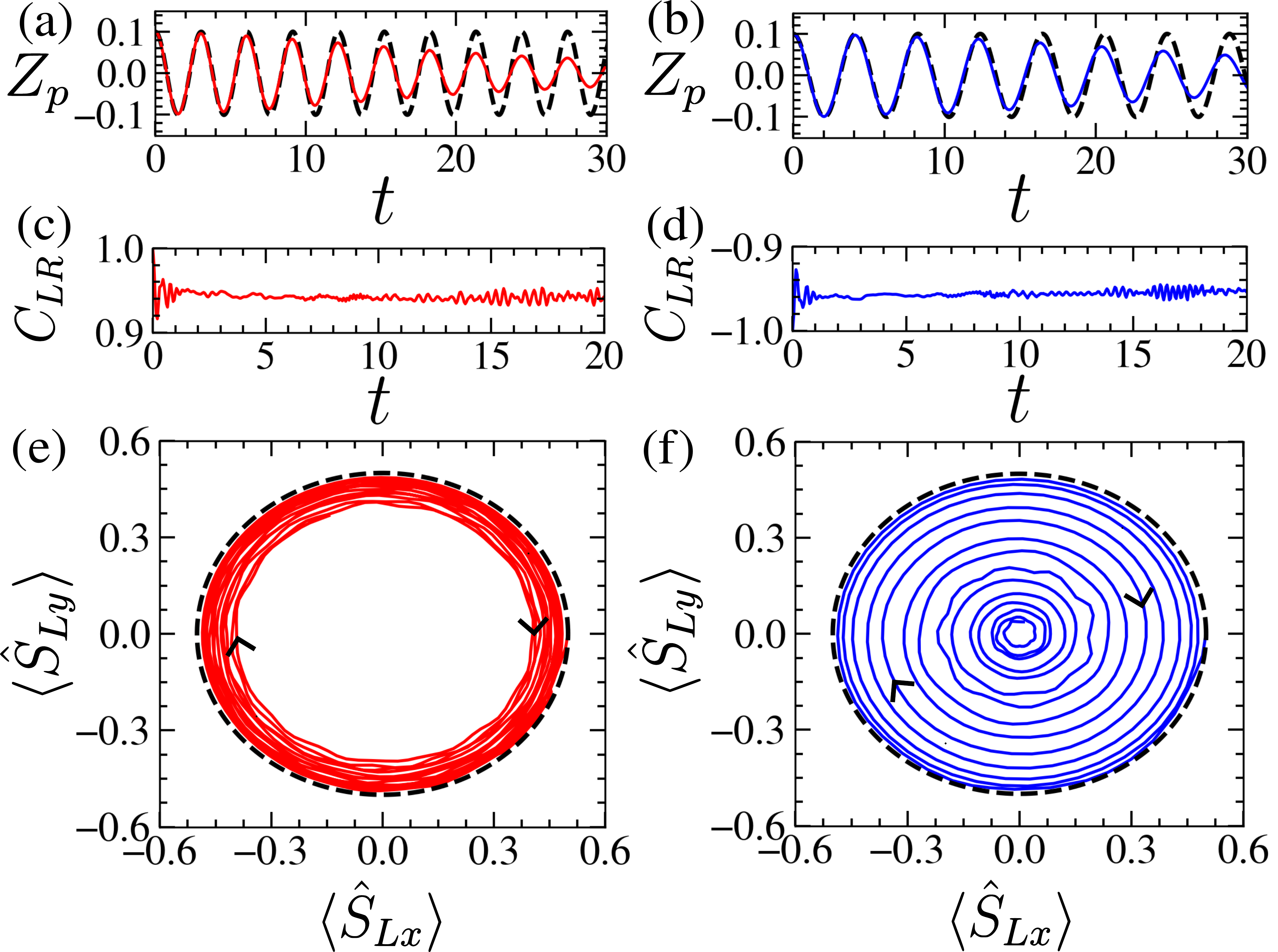}
	\caption{{\it Comparison between quantum dynamics of the symmetric states FP-F and FP-AF:} The small amplitude oscillation of the photon imbalance $Z_p$ around (a) FP-F, (b) FP-AF, are compared with their respective classical dynamics (black dashed lines). The spin orientations of these states are described by $C_{LR}$ in (c,d) respectively. The spin trajectory in the $\langle \hat{S}_{Lx}\rangle$-$\langle \hat{S}_{Ly}\rangle$ plane of the left cavity is shown for (e) FP-F and (f) FP-AF state. The black dashed circles in (e,f) represent the ring of classical FPs. The evolution of spins around the FP-AF state exhibits dephasing phenomenon. Parameter chosen: ($\tilde{g},\tilde{U}$)=($0.5,0.5$). }
	\label{fig7}
\end{figure}

First, we study the dynamics of the symmetric states FP-F and FP-AF corresponding to the ferromagnetic and the anti-ferromagnetic classes respectively. 
To characterize these states quantum mechanically, we obtain the photon population imbalance, $Z_p=(\langle \hat{n}_L\rangle-\langle \hat{n}_R\rangle)/(\langle \hat{n}_L\rangle+\langle \hat{n}_R\rangle)$ where $\langle \hat{n}_i\rangle$ is computed from the time evolved state $\ket{\Psi(t)}$ starting from the initial coherent state. 
For both FP-F and FP-AF states, we obtain the time evolution of $Z_p$ and compare them with that obtained from the classical dynamics, as shown in Fig.\ref{fig7}(a,b). 
It is clear from Fig.\ref{fig7}(a,b) that the simple classical analysis is able to capture the full quantum dynamics reasonably well, however, there are certain deviations as $t$ increases. 
To reveal the relative spin orientation in the two cavities, we introduce the quantity
\begin{eqnarray}
C_{LR} = \frac{\langle \hat{S}_{Lx}\hat{S}_{Rx}+\hat{S}_{Ly}\hat{S}_{Ry}\rangle}{\sqrt{\left(\frac{1}{4}-\langle \hat{S}_{Lz}\rangle^2\right)\left(\frac{1}{4}-\langle \hat{S}_{Rz}\rangle^2\right)}},
\label{Spin-orientation}
\end{eqnarray}
which in the classical limit takes the value -1(+1) corresponding to the (anti)ferromagnetic class of steady states. As shown in Fig.\ref{fig7}(c,d), the quantum dynamics of $C_{LR}$ also approaches these values for FP-F and FP-AF, which is consistent with their classification based on classical analysis.  
On the other hand, in quantum dynamics, the correlation (entanglement) between spins and photons gives rise to interesting effects leading to the deviation from classicality. In the spin dynamics of the FP-F state, the average values of the spin components in the $x$-$y$ plane evolve around a circle corresponding to the classical FPs. Whereas for the FP-AF state, the spin trajectory deviates from the ring of classical FPs and spirals to the center corresponding to $\langle \hat{S}_x\rangle = \langle \hat{S}_y\rangle =0$, exhibiting spin dephasing phenomena \cite{dephasing}, as seen from Fig.\ref{fig7}(f).
Typically for spin $1/2$ qubits, the classical description fails due to the enhanced quantum fluctuations and entanglement with photons, which we analyze later.
\begin{figure}
	\centering
	\includegraphics[width=\columnwidth]{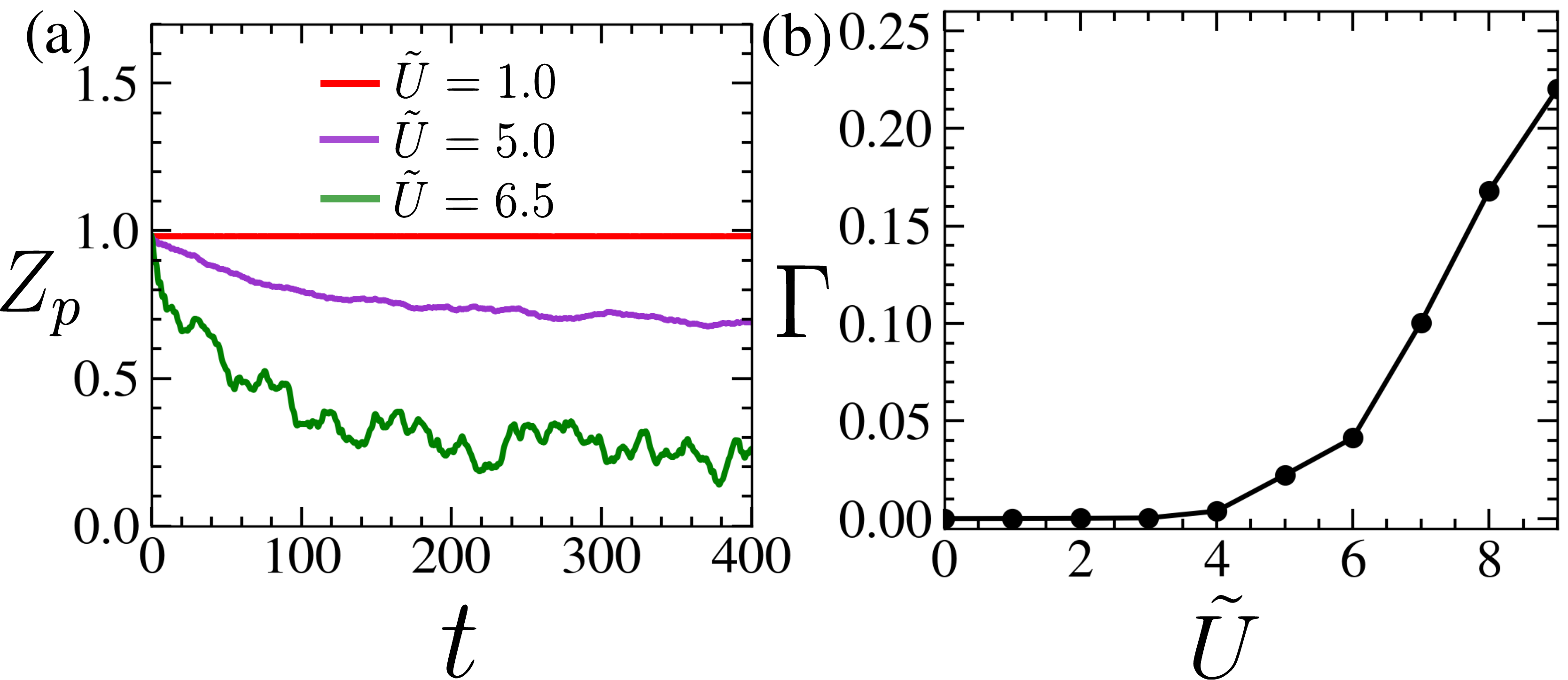}
	\caption{\textit{Quantum dynamics corresponding to the perfect self trapped state PST:}  (a) Time evolution of the relative photon population imbalance $Z_p$ for $\tilde{g}=5$ and different Kerr nonlinearity $\tilde{U}$. The imbalance decays above a certain large value of Kerr interaction strength. (b) Variation of the decay rate $\Gamma$ of the imbalance with $\tilde{U}$. 
	}
	\label{fig8}
\end{figure}

Next, we investigate different types of self-trapping phenomena from quantum dynamics. We search for a perfect self trapped state, where almost all the photons become localized in one of the cavities. It is evident from the classical phase diagram that atom-photon interaction is crucial for perfect self-trapping of photons. For small Kerr nonlinearity, we identify the perfect self trapped state quantum mechanically, for which the relative imbalance of photon $Z_p$ remains close to unity for a sufficiently long time (see the red line in Fig.\ref{fig8}(a)). 
\begin{figure}[b]
	\centering
	\includegraphics[width=\columnwidth]{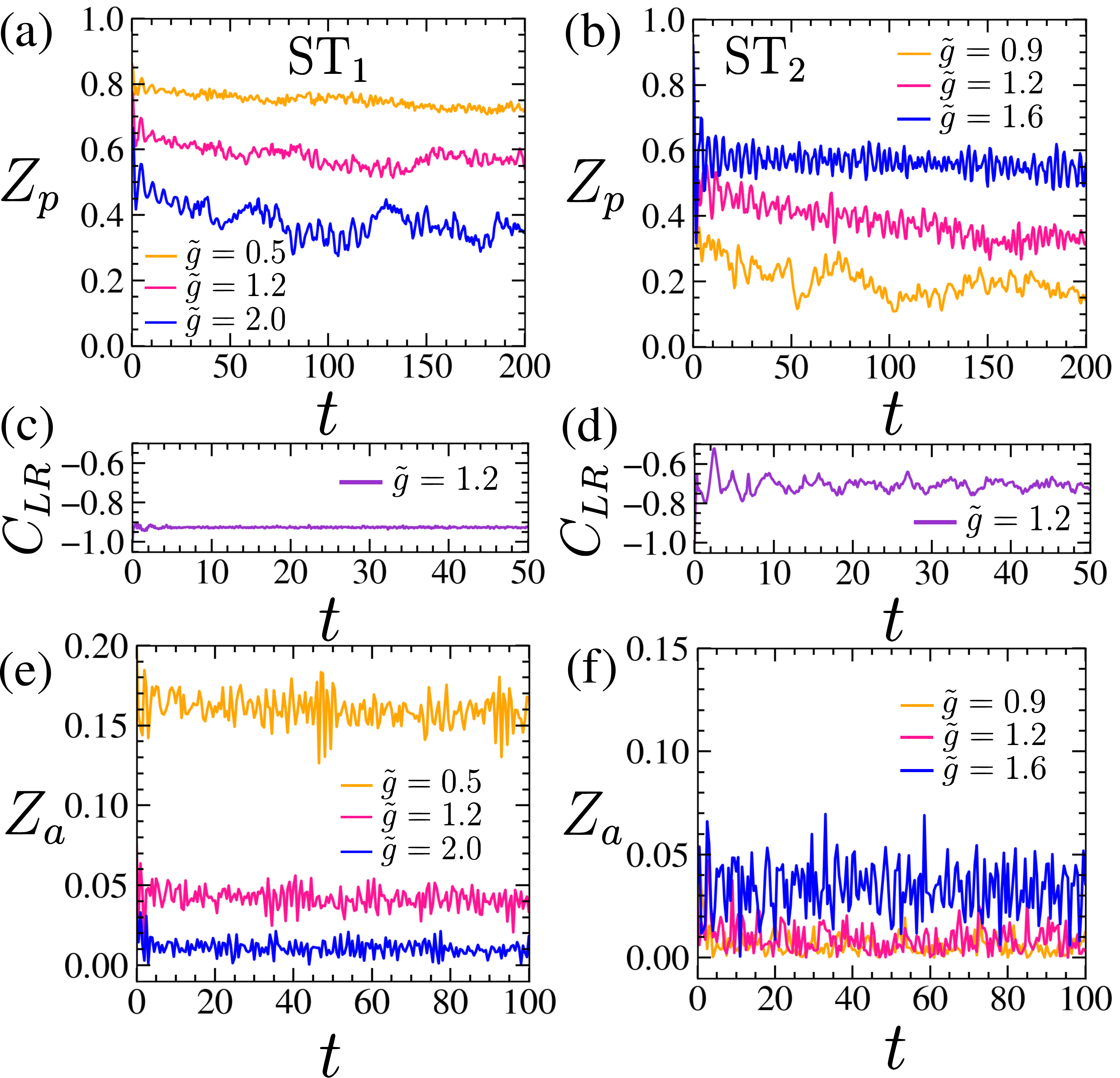}
	\caption{{\it Comparison between the quantum dynamics of the two distinct self trapped states ST$_1$ and ST$_2$:}  The plots (a,c,e) and (b,d,f) correspond to dynamics of various quantities around ST$_1$ and ST$_2$ states for $\tilde{U}=5$ and $\tilde{U}=2$, respectively. Time evolution of the (a,b) relative photon imbalance $Z_p$ and (e,f) atomic imbalance $Z_a$, for different coupling strength $\tilde{g}$. (c,d) Dynamics of the spin correlation function $C_{LR}$ at $\tilde{g}=1.2$.
	}
	\label{fig9}
\end{figure}
However, for sufficiently large Kerr nonlinearity, the imbalance becomes significantly lower than unity and decays with time, which is shown in Fig.\ref{fig8}(a). The rate of exponential decay $\Gamma$ can be obtained by numerically fitting the time evolution of the imbalance. The variation of the decay rate with Kerr nonlinearity exhibits an interesting feature as it grows rapidly above certain Kerr nonlinearity, which is depicted in Fig.\ref{fig8}(b). This indicates that sufficiently large Kerr nonlinearity induces an instability in perfect self-trapping. Although, classically, the stable perfect self trapped state exists for $\tilde{g}>\tilde{g}_{c1}$, the sufficiently large Kerr nonlinearity gives rise to instability in such a state during quantum dynamics. 

We also analyze the self trapped states ST$_1$ and ST$_2$ of the anti-ferromagnetic class, shown in Fig.\ref{fig9}(a,b). As a characteristic feature of these states, we study the dynamics of the photon imbalance $Z_p$ for increasing values of atom-photon coupling strength $\tilde{g}$. As depicted in Fig.\ref{fig9}(a), for ST$_1$, the imbalance decreases with $\tilde{g}$, whereas, it increases for ST$_2$ (see Fig.\ref{fig9}(b)), which is consistent with the classical analysis and can be used to distinguish between these two self trapped states. Since both these states belong to the anti-ferromagnetic class, during dynamics, the quantity $C_{LR}$ acquires a negative value for both states, however, its deviation from the classical value is large for ST$_2$ (depicted in Fig.\ref{fig9}(c,d)). 
We also study the dynamics of the atomic imbalance $Z_a=|\langle\hat{S}_{Lz}\rangle-\langle\hat{S}_{Rz}\rangle|$, which decreases with increasing coupling strength $\tilde{g}$ for ST$_1$, as shown in Fig.\ref{fig9}(e), that is in agreement with the classical analysis (see Fig.\ref{fig6}). For the ST$_2$ state, the evolution of $Z_a$ always saturates to a very small value, exhibiting a weak variation with $\tilde{g}$ (see Fig.\ref{fig9}(f)), which is in stark contrast with the ST$_1$ state.
The above analysis reveals that the deviation from classicality is significantly large for the ST$_2$ state as compared to ST$_1$. In spite of such quantitative differences, the main features of the semiclassical steady states are observed in the quantum dynamics.

It is worth mentioning that new effects in dynamics can arise, since dissipation is inevitable in such systems, specifically due to photon loss and spontaneous emission from atoms, which can however be controlled by the appropriate experimental setup \cite{Dissipation1,Dissipation2,Dissipation3,Dissipation4}. 		
For weak dissipation, it is expected that in JCJJ the photonic oscillations corresponding to the symmetric states will be damped, nevertheless, their signature can be detected from the oscillation frequency.
Similarly, the self-trapping phenomenon can still be detected within its lifetime due to a decay in the photonic imbalance.
Even in the presence of dissipation, the self trapping phenomenon has already been observed in different experimental setups \cite{Self_trapping1,JC_dimer_expt,BJJ_self_trapped}.
It is possible to stabilize the steady states through the incoherent pumping processes, balancing the photon loss \cite{Dissipative_transition7,JCD_Manus_K}.

In addition to the characteristic features of the steady states, it is crucial to investigate other quantum effects manifested in the dynamics. These include the generation of entanglement between the qubit and the photonic degree of freedom, as well as the quantum state of the photon field, which we discuss in the next section.

\section{ENTANGLEMENT AND QUANTUM FLUCTUATIONS}\label{5}
The semiclassical formalism presented in Sec.\ref{3} is based on the product coherent state representation, which is appropriate for describing the phase coherent photonic Josephson dynamics. 
However, the presence of interactions and Kerr nonlinearity can destroy such coherent dynamics due to enhanced phase fluctuations, which in turn gives rise to the deviation from classicality due to a change in the nature of the quantum state.  
To this end, we study the phase fluctuations of the photon field by constructing the phase states \cite{Oberthalar,phaseBarnett},
\begin{eqnarray}
	\ket{\psi_m}=\frac{1}{\sqrt{N_{max}+1}}\sum_{n=0}^{N_{max}}\exp(in\psi_m)\ket{n}
\end{eqnarray}
with $\psi_m=\psi_0+2\pi m/(N_{max}+1)$, where $m$ is an integer $m\in [0,N_{max}]$ and  $\psi_m\in [-\pi,\pi]$. These phase states are eigenstates of the phase operator $\hat{\psi}$, as given by,
\begin{equation}
		\exp(\pm i\hat{\psi})\ket{\psi_m}=\exp(\pm i\psi_m)\ket{\psi_m}.
\end{equation}
The phase distribution corresponding to the photon field in one of the cavities ($i = L,R$) is given by,
\begin{eqnarray}
		{\rm P}(\psi_m^i) = {\rm Tr}(\hat{\rho}^{i}_{p}\ket{\psi_{m}}\bra{\psi_{m}})
\end{eqnarray}
with $\sum_{m}{\rm P}(\psi_m^i)=1$, where $\hat{\rho}_{p}^i$ is the reduced density matrix corresponding to the photon field of the $i$th cavity, obtained by tracing out the other degrees. The average value and the fluctuation of the phase of the photon field in each cavity can be computed from the phase distribution as,
\begin{subequations}
		\begin{eqnarray}
			\langle\hat{\psi}_i\rangle &=&\sum_m\psi_m {\rm P}(\psi^i_m)\\
			(\Delta\hat{\psi}_i)^2&=&\sum_m(\psi_m-\langle\hat{\psi}_i\rangle)^2\,{\rm P}(\psi^i_m).
		\end{eqnarray} 
\end{subequations} 	
Using the above prescription, we compute the mean phase difference between the cavity modes $\psi_r=\langle\hat{\psi}_L\rangle-\langle\hat{\psi}_R\rangle$ and its time evolution.
\begin{figure}
	\centering
	\includegraphics[width=\columnwidth]{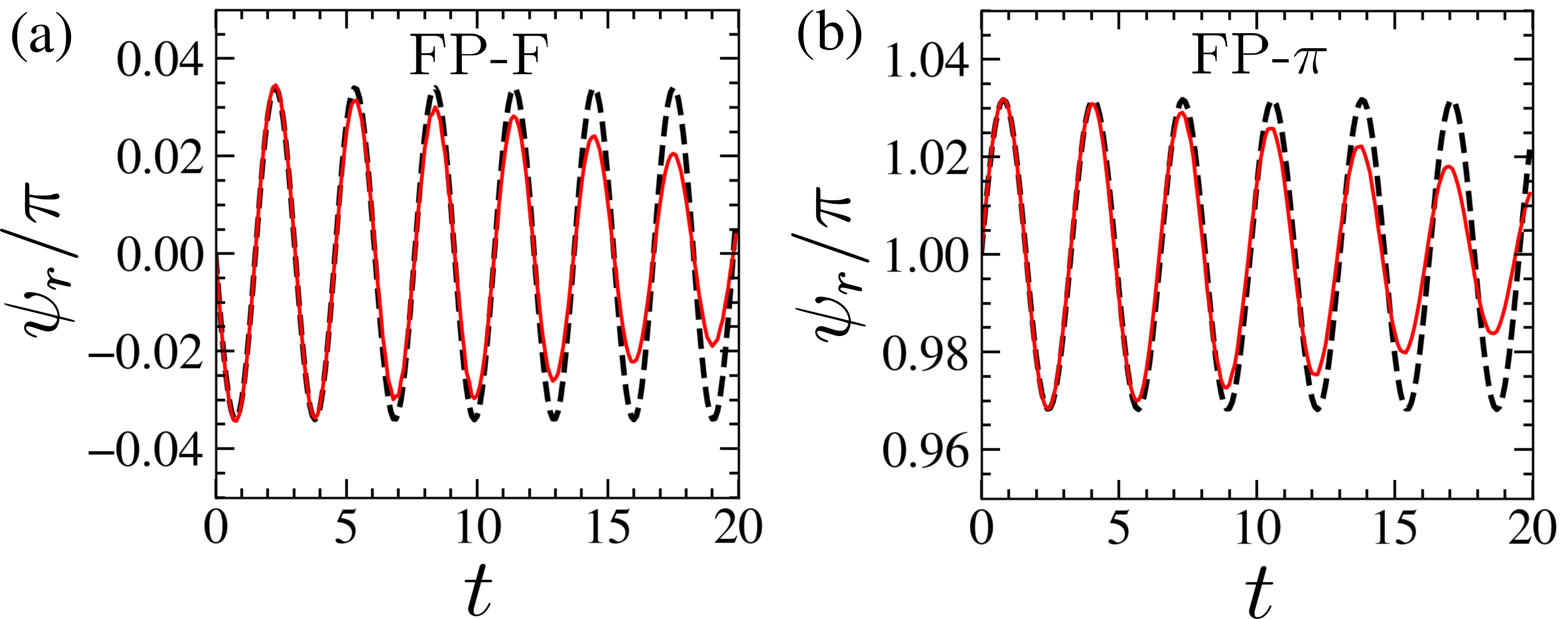}
	\caption{\textit{Classical-quantum correspondence for the symmetric states FP-F and FP-$\pi$:}  Quantum dynamics of the small amplitude oscillation of the relative phase $\psi_r = \langle \hat{\psi}_L\rangle-\langle \hat{\psi}_R\rangle$ of the photon field corresponding to (a) FP-F and (b) FP-$\pi$ states for the coupling strengths ($\tilde{g}$,$\tilde{U}$)=(0.5,0.5). The black dashed lines represent the classical dynamics around these states.}
	\label{fig10}
\end{figure}
The dynamics of the relative phase of the photon modes for the symmetric FP-F and FP-$\pi$ states are shown in Fig.\ref{fig10}(a) and (b) respectively, which exhibit coherent oscillations around their steady state values $0$ and $\pi$. 
\begin{figure}[b]
	\centering
	\includegraphics[width=\columnwidth]{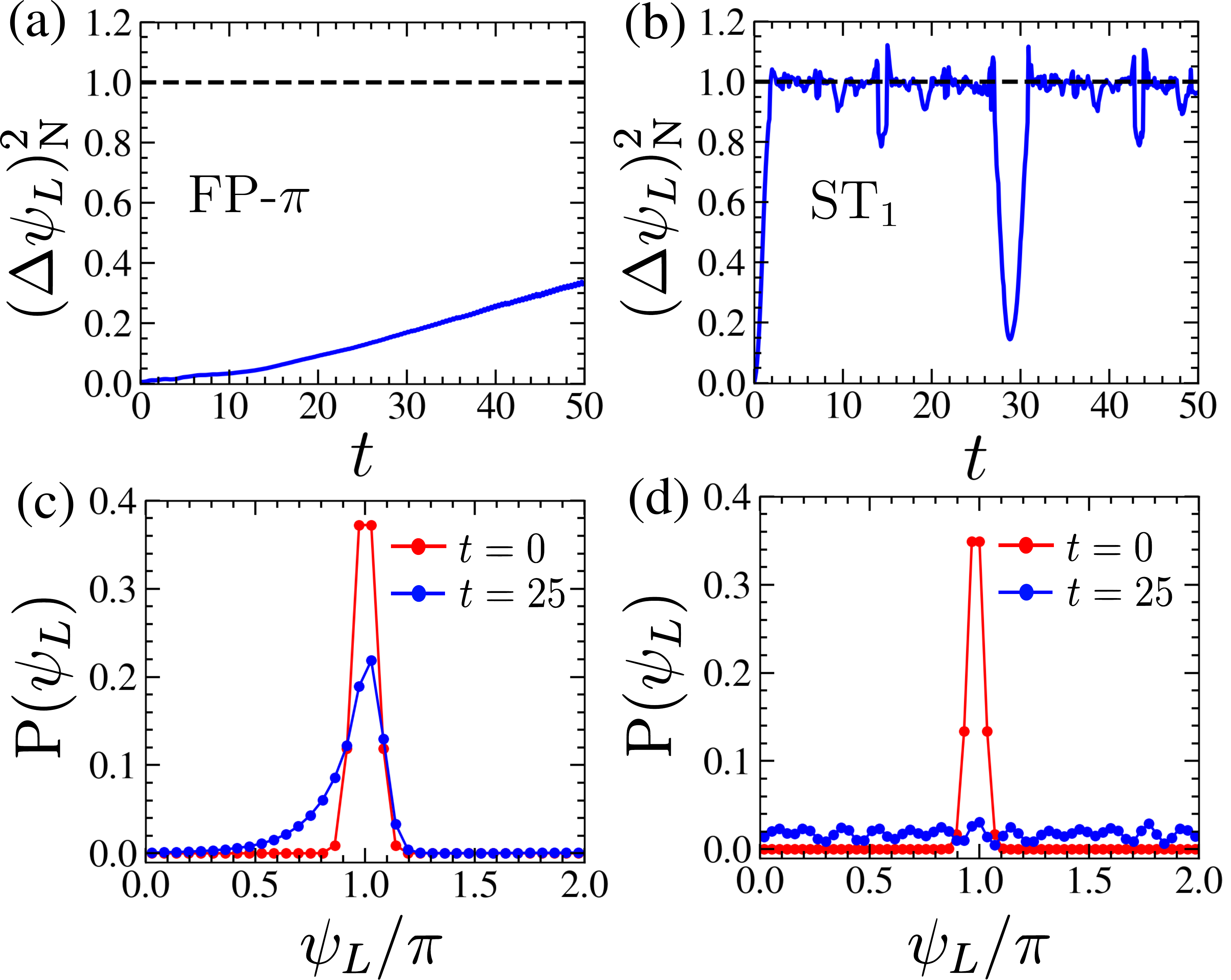}
	\caption{{\it Phase diffusion dynamics:} Time evolution of the relative phase fluctuation $(\Delta \psi)^2_{\rm N}$ for (a) symmetric state FP-$\pi$ with $(\tilde{g},\tilde{U})=(1.0,0.5)$ and (b) self trapped state ST$_1$ for $(\tilde{g},\tilde{U})=(1.0,6.5)$. The horizontal dashed line indicates the maximum value $(\Delta \psi)^2_{\rm N}=1$. (c,d) Snapshots of the corresponding phase distributions at different times.}
	\label{fig11}
\end{figure}
To quantify the degree of coherence, we calculate the normalized phase fluctuation of photons $(\Delta\psi_i)^2_{\rm N}=(\Delta\psi_i)^2/(\Delta\psi_i)^2_{\rm max}$ in one of the cavities ($i=L,R$) where the maximum phase fluctuation $(\Delta\psi_i)^2_{\rm max}=\pi^2/3$ corresponds to a uniform phase distribution \cite{Oberthalar}.
For the FP-$\pi$ state, the phase fluctuation remains small during time evolution, as illustrated in Fig.\ref{fig11}(a), due to which the coherent phase oscillation is retained. 
However, the phase fluctuation of the FP-$\pi$ state increases slowly, and after a sufficiently long time, it approaches the maximum value.
On the other hand, an enhancement in the growth of phase fluctuation can be observed for the self trapped state ST$_1$, arising for large $\tilde{U}$, as seen from Fig.\ref{fig11}(b).
In general, the phase fluctuation increases with Kerr nonlinearity, which is evident from the above comparison.
Such enhanced phase fluctuation during the time evolution is associated with the broadening of the phase distribution, indicating the deviation of the photon field from its classical representation in terms of the coherent state.
Spreading of the phase distribution of FP-$\pi$ and ST$_1$ states during time evolution is apparent from Fig.\ref{fig11}(c,d).  
Even though the phase fluctuation attains its maximum value almost immediately for the ST$_1$ state, the appearance of dips in the time evolution of $(\Delta\psi_L)^2_{\rm N}$, as observed from Fig.\ref{fig11}(b), corresponds to the revival of the phase of the photon field, which we discuss later. 
\begin{figure}
	\centering
	\includegraphics[width=\columnwidth]{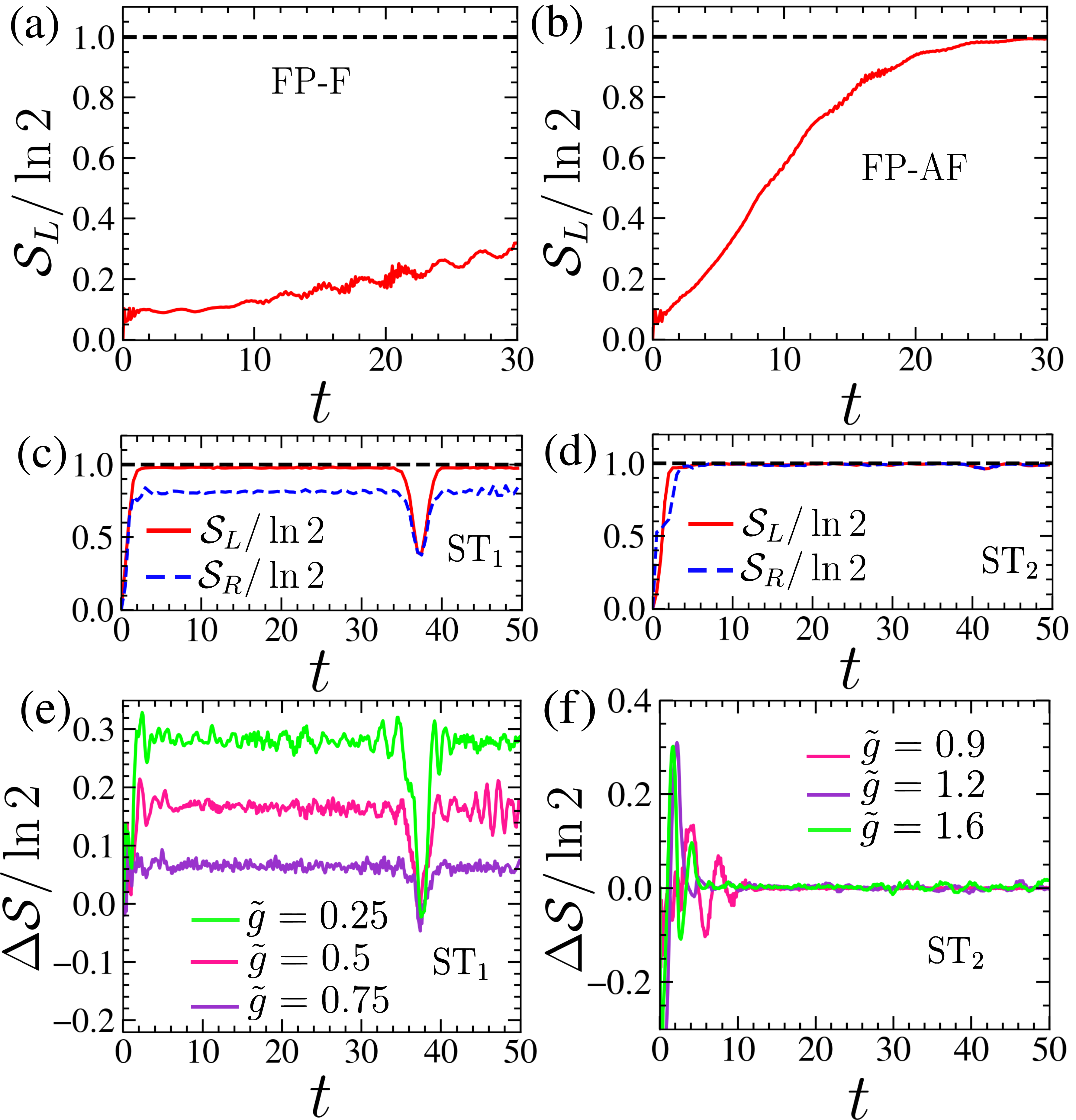}
	\caption{{\it Entanglement entropy of the spin degree for various dynamical states:} Dynamics of the scaled entanglement entropy $\mathcal{S}_L$ of the spin in the left cavity corresponding to (a) FP-F and (b) FP-AF for $(\tilde{g},\tilde{U})=(0.5,0.5)$. Entanglement entropy of the spins in both the cavities corresponding to (c) ST$_1$ state for $(\tilde{g},\tilde{U})=(0.5,5.0)$ and (d) ST$_2$ state for $(\tilde{g},\tilde{U})=(1.2,2.0)$. (e,f) Variation of the relative entanglement entropy $\Delta \mathcal{S}=\mathcal{S}_L-\mathcal{S}_R$ with coupling strength $\tilde{g}$ for ST$_1$ and ST$_2$ states, respectively. The Kerr nonlinearity in (e,f) is chosen the same as that in (c,d).
	}
	\label{fig12}
\end{figure}

Apart from the phase fluctuation, the entanglement between the photon field and spins during the time evolution gives rise to interesting quantum effects and deviation from classicality. 
Starting from the total density matrix $\hat{\rho} = \ket{\Psi(t)}\bra{\Psi(t)}$, computed from the full wavefunction $\ket{\Psi(t)}$, the reduced density matrix of a subsystem (such as the spin/photon field of each cavity) can be obtained by integrating out the rest of the degrees of freedom. Following this prescription, we compute the entanglement entropy of the subsystem (corresponding to the cavities) as,
\begin{equation}
\mathcal{S}_{i}=-\sum_{l}\lambda_{l}^i\log(\lambda_{l}^i)
\end{equation} 
where, $\lambda_l^i$ represents the eigenvalue with index $l$ of the reduced density matrix corresponding to the subsystem denoted by $i$ (for example, $i=L,R$ is the cavity index). In a similar manner, we can also compute the reduced density matrix and entanglement entropy $\mathcal{S}_{LR}$ for the total photon and spin degree separately.
Ideally, $\mathcal{S}_{i}$ vanishes for product state, but due to atom-photon interactions, the entanglement entropy increases during time evolution.
We obtain the entanglement entropy $\mathcal{S}_i$ of the spin in each cavity, corresponding to the symmetric states FP-F and FP-AF, which are compared in Fig.\ref{fig12}(a,b). Unlike the FP-F state, $\mathcal{S}_i$  grows rapidly and saturates to its maximum value $k_{B}\ln2$ for the FP-AF state, due to which the spin dynamics deviates from the classical steady states, exhibiting dephasing phenomenon (see Fig.\ref{fig7}(f)), as discussed in Sec.\ref{4}.
\begin{figure}[t]
	\centering
	\includegraphics[width=\columnwidth]{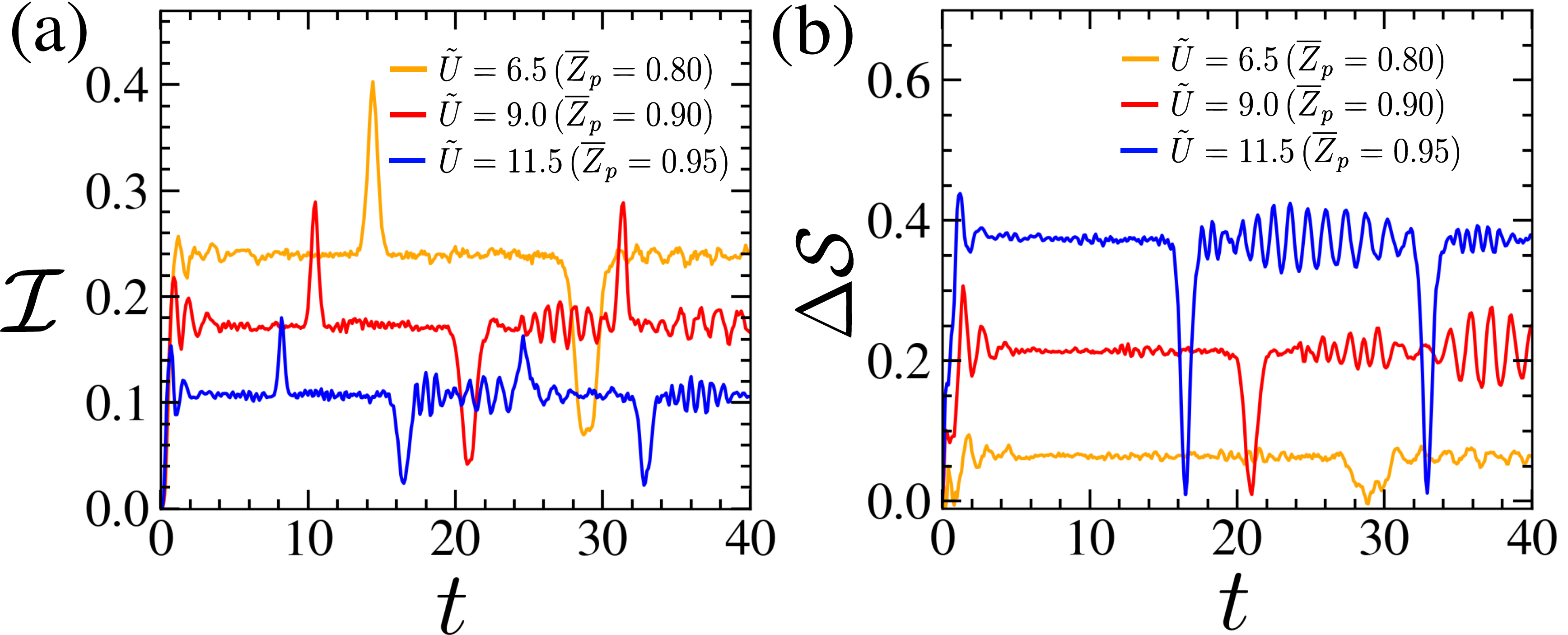}
	\caption{{\it Photon mediated correlation between the spins in the different cavities for the ST$_1$ state:} The dynamics of (a) mutual information $\mathcal{I}$ and (b) relative entanglement entropy $\Delta \mathcal{S}$ between the spins in the cavities for different Kerr strengths $\tilde{U}$. The saturation value of the photon population imbalance $\overline{Z}_p$ obtained after long time, corresponding to these values of $\tilde{U}$ are mentioned in the figure. Here, we set $\tilde{g} = 1$.}
	\label{fig13}
\end{figure}

We also compare the entanglement entropy of spins in both the cavities for the self trapped states ST$_1$ and ST$_2$, which reveals contrasting features between them.
For ST$_2$, the $\mathcal{S}_{i}$ is almost the same for both the cavities and saturates to the maximum value. On the contrary, for ST$_1$, the entanglement entropy is larger, corresponding to the cavity containing more number of photons, as seen from Fig.\ref{fig12}(c,d).
In addition, we also study the difference between the entanglement entropy of spins in the two cavities $\Delta \mathcal{S} = \mathcal{S}_L-\mathcal{S}_R$ and their variation with coupling strength $\tilde{g}$, as shown in Fig.\ref{fig12}(e,f). For ST$_1$ state, similar to the photon imbalance $Z_p$, the saturation value of $\Delta \mathcal{S}$ decreases with increasing $\tilde{g}$, which is in stark contrast to ST$_2$ state, for which $\Delta \mathcal{S}$ vanishes over time, irrespective of the values of $\tilde{g}$. However, the timescale at which $\Delta \mathcal{S}$ vanishes exhibits a weak variation with $\tilde{g}$ within a small range.  
Such contrasting feature of entanglement dynamics of two qubits can also distinguish the self trapped states ST$_1$ and ST$_2$. 

Apart from the interaction induced entanglement between spins and photons in each cavity, two apparently non interacting spins can also be entangled, which is mediated by photons.  Such photon induced hidden correlation between two spins can be analyzed from the mutual information \cite{Nilson_Chuang,MI1,MI2,MI3,MI4,MI5},
\begin{equation}
	\mathcal{I}=\mathcal{S}_{L}+\mathcal{S}_{R}-\mathcal{S}_{LR},
\end{equation}  
which reveals very interesting behavior for the self trapped states. For the ST$_2$ state, both $\Delta \mathcal{S}$ and $\mathcal{I}$ are very small, exhibiting almost no variation with interaction strengths, which indicates that the reduced density matrix corresponding to the two spins approaches to the maximally mixed state \cite{Nilson_Chuang}.
On the other hand, in the case of ST$_1$, increasing the photon population imbalance leads to an increase in $\Delta \mathcal{S}$, while the mutual information $\mathcal{I}$ decreases, as evident from Fig.\ref{fig13}. 
\begin{figure}[b]
	\centering
	\includegraphics[width=\columnwidth]{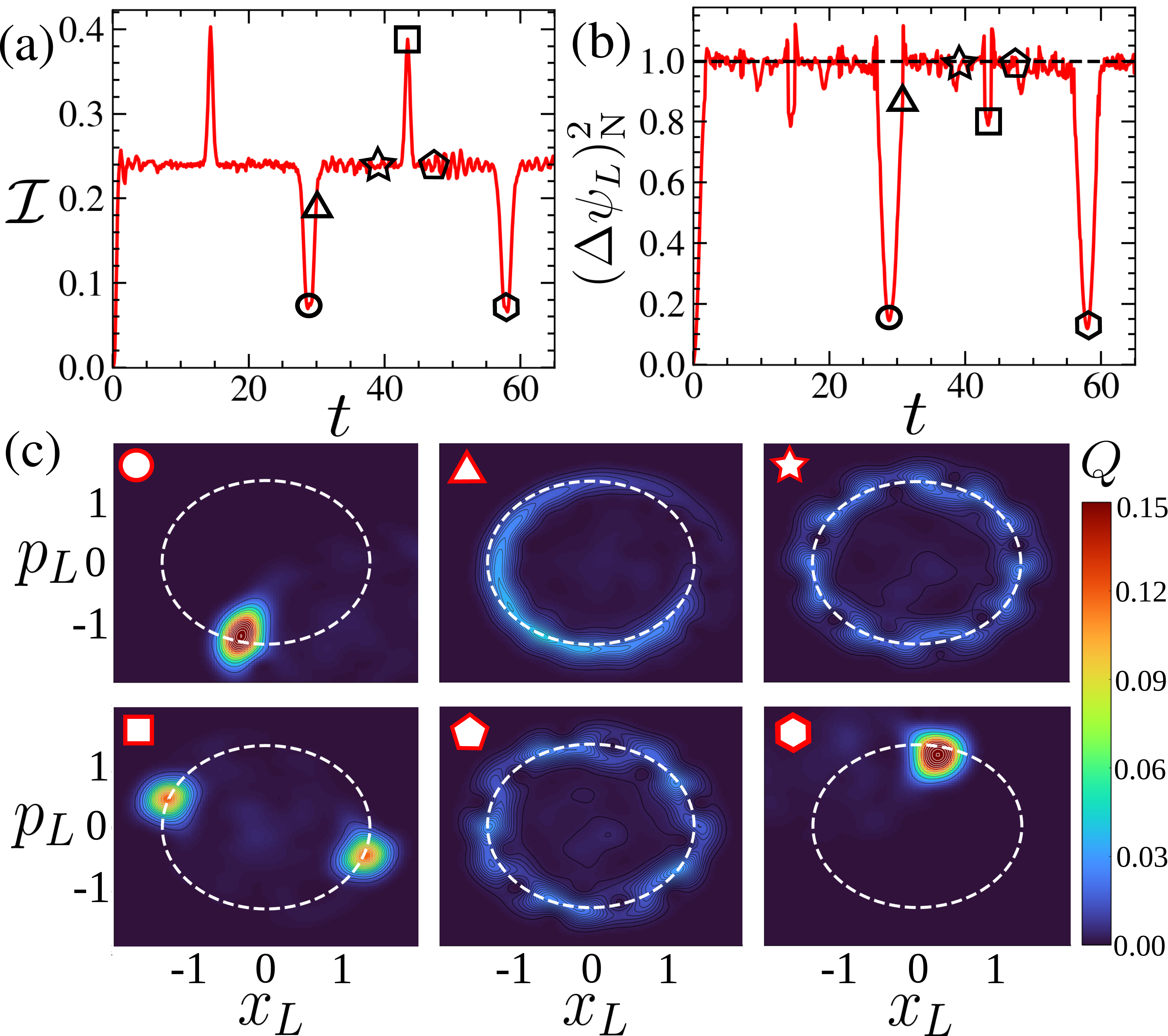}
	\caption{\textit{Evolution of the quantum state and its signature in the phase fluctuation and mutual information}:  Dynamics of (a) mutual information $\mathcal{I}$ of the spins and (b) relative phase fluctuation $(\Delta \psi_L)^2_{\rm N}$ of the photon field, corresponding to ST$_1$ state. (c) Husimi distribution $Q$ in the $x_L$-$p_L$ plane of the photon field at different time instances of the revival cycle, marked in (a,b) by different symbols. The white dashed lines in (c) represent the ring of FPs corresponding to the ST$_1$ state. Parameter chosen: $(\tilde{g}, \tilde{U})=(1.0,6.5)$.}
	\label{fig14}
\end{figure}
Such tunability of quantum correlations between the two non interacting spins in the cavities can have potential applications in quantum information processing. Additionally, for the ST$_1$ state, the mutual information $\mathcal{I}$ exhibits dip and spike like structure during the time evolution, as seen from Fig.\ref{fig14}(a). Such dips in the mutual information  correspond to the phase revival phenomenon \cite{Revival_1,Revival_2,Revival_3,revival}, resulting in a sudden drop in phase fluctuation, as seen in Fig.\ref{fig14}(b). This revival cycle can be analyzed from the evolution of the semiclassical phase space density of the photon field, described by the Husimi distribution,
\begin{equation}
	Q(\alpha,\alpha^*)=\frac{1}{\pi}\bra{\alpha}\hat{\rho}_i^p\ket{\alpha},
\end{equation} 
where $\hat{\rho}_i^p$ represents the reduced density matrix of the photon field in the cavities. 
In order to compare the Husimi distribution with the corresponding classical dynamics, we introduce the scaled phase space variables $x_i=(\alpha_i+\alpha_i^*)/\sqrt{2M}$ and $p_i=(\alpha_i-\alpha_i^*)/i\sqrt{2M}$, where 
$M$ represents the conserved total excitation number associated with the initial state. 
Initially, the density is localized around one of the FPs, exhibiting the  coherent structure of the photon field. As time evolves, the phase space density spreads over the ring of fixed points, describing the loss of coherence and finally, it is reconstructed at a point in the phase space when another dip in $(\Delta \psi)^2_{\rm N}$ occurs, exhibiting the revival phenomenon (see Fig.\ref{fig14}(c)). Interestingly, in the middle of the cycle, the phase space density splits and becomes localized around two diagonally opposite phase space points. Although such bimodal phase space distribution of photons resembles that of a cat state \cite{Cat_1,Cat_2,Cat_3,Cat_4,Cat_5,Dan_Walls}, they differ in terms of their coherence property, which can be captured from the Wigner function, as discussed in Appendix.\ref{Appendix_a}.
Such structure of phase space density is associated with the appearance of a spike in the mutual information, as seen from Fig.\ref{fig14}(a).

The above analysis elucidates fascinating quantum effects and entanglement associated with the evolution of the quantum state, corresponding to different dynamical branches, which can also be relevant in the context of quantum information processing.
Such quantum effects give rise to the deviation from classicality, nevertheless, the qualitative behavior of the system can still be captured from the coherent state description. 
Apart from the steady state dynamics, it would also be interesting to investigate the evolution of the quantum state, particularly that of the photon field, when the system is driven to the unstable regime.

\paragraph*{\it Quench dynamics to unstable regime:} 

\begin{figure}
	\centering
	\includegraphics[width=\columnwidth]{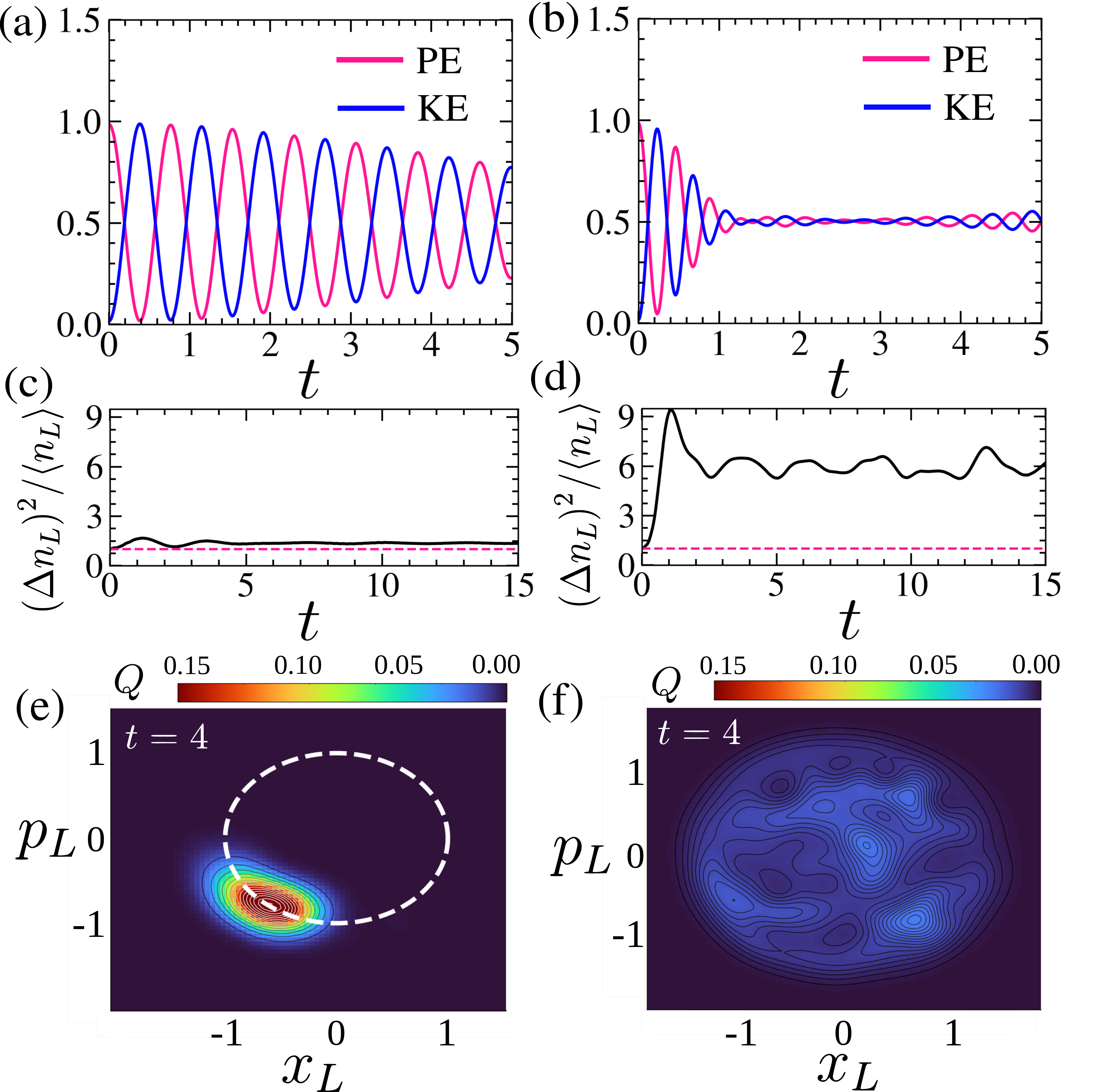}
	\caption{\textit{Quench dynamics under sudden change in the Kerr nonlinearity $\tilde{U}$:} The initial state is prepared corresponding to the stable FP-$\pi$ state for $\tilde{U}=0.5$ before the quench. (a,c,e) and (b,d,f) correspond to the dynamics after a quench to the stable regime of FP-$\pi$ with $\tilde{U}=1.5$ and unstable regime at $\tilde{U}=6.5$, respectively.
	(a,b) the time evolution of the scaled potential and kinetic energies, PE $ = m\omega \langle \hat{x}_L^2\rangle/2\hbar M$ (red) and KE $ =\langle \hat{p}_L^2\rangle/2\hbar m\omega M$ (blue), respectively. (c,d) depicts the dynamics of scaled photon number variance $(\Delta n_L)^2/\langle n_L\rangle$. The pink dashed line in (c,d) represents $(\Delta n_L)^2/\langle n_L\rangle=1$, corresponding to the coherent state. (e,f) Husimi distribution $Q$ of the photon field in the left cavity at time $t=4$. The white dashed line in (e) represents the ring of steady states corresponding to the FP-$\pi$ state. Parameter chosen: $\tilde{g}=0.5$.
	}
	\label{fig15}
\end{figure}	
Next, we investigate the quench dynamics corresponding to an abrupt change in the Kerr nonlinearity $\tilde{U}$, starting from the initial coherent state corresponding to the stable FP-$\pi$ state.
First, we consider a small change in $\tilde{U}$, for which the FP-$\pi$ state remains stable. Under this sudden change, the system still follows the stable FP-$\pi$ branch, exhibiting oscillations around it. 
After quench, the initial coherent state begins to move around the ring of FPs, as depicted in Fig.\ref{fig15}(e). During the time evolution, the wavefunction initially remains fairly localized and then slowly spreads along the ring of FPs.
As a consequence, the scaled potential energy (PE) and kinetic energy (KE) $\left(m\omega \langle \hat{x}_i^2\rangle/2\hbar M,\langle \hat{p}_i^2\rangle/2\hbar m\omega M\right)$ of the photon field in each cavity oscillate coherently, keeping the average photon number fixed (see Fig.\ref{fig15}(a)). Moreover, the photon phase fluctuation increases slowly and finally saturates to its maximum value after a sufficiently long time, while its phase space density remains localized around the ring of FPs. 

\begin{figure}[t]
	\centering
	\includegraphics[width=\columnwidth]{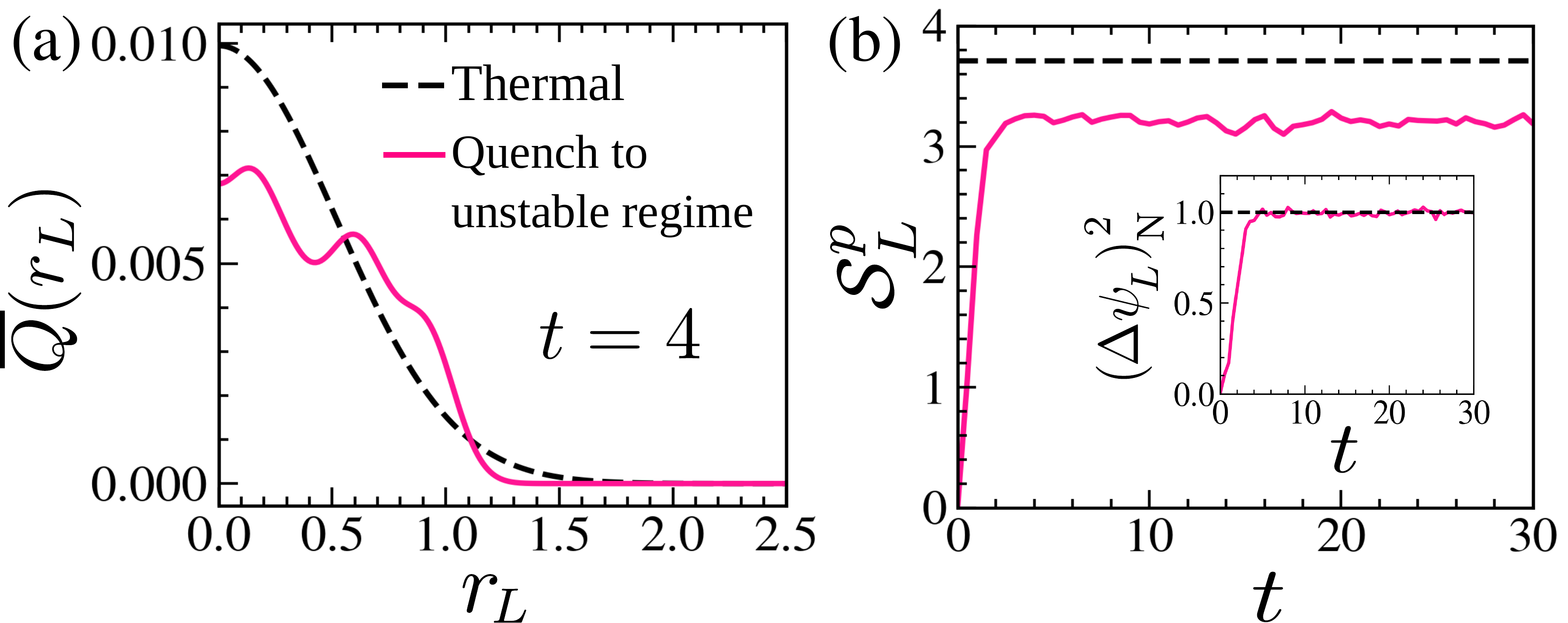}
	\caption{{\it Comparison of the photon field with the thermal state, after the quench to the unstable regime.} (a) Angular averaged Husimi distribution of the photon field in the left cavity  $\overline{Q}(r_L)$ (solid red line) compared with the thermal distribution (black dashed line) at time $t=4$. (b) Dynamics of the entanglement entropy $\mathcal{S}_L^p$ of the photon field in the left cavity. The entropy of the thermal state is marked by the black dashed line. The inset of (b) shows the time evolution of the relative phase fluctuation $(\Delta \psi)^2_{\rm N}$, which saturates to unity. }
	\label{fig16}
\end{figure}

On the contrary, when the interaction strength $\tilde{U}$ is quenched above the transition point where the $\pi$-mode becomes unstable, the system exhibits incoherent dynamics, dominated by large fluctuations, instead of following any stable branch of the self trapped state. 
After quenching, the photon field loses its coherence rapidly, as the phase fluctuation attains the maximum value (See inset of Fig.\ref{fig16}(b)). Moreover, the scaled kinetic and potential energies approach the same steady value, without exhibiting large amplitude oscillations (see Fig.\ref{fig15}(b)), analogous to the equipartitioning of energy. In this regime, the reduced density matrix of the photon field in each cavity exhibits a predominant contribution from the diagonal elements. Simultaneously, the associated Husimi distribution spreads widely across the phase space, revealing substantial fluctuations in photon number $(\Delta n_i)^2=\langle \hat{n}_i^2\rangle-\langle \hat{n}_i\rangle^2\gg \langle n_i\rangle$, notably in contrast to the initial coherent state for which $(\Delta n_i)^2 = \langle n_i\rangle$. 
The entanglement entropy of the photon field grows linearly with time and finally saturates, as seen from Fig.\ref{fig16}(b). Moreover, the reduced density matrix of the spins in two cavities approaches the maximally mixed state.
Such scenarios of quench dynamics to an unstable regime resemble thermalization, which led us to compare the state of the photon field with that of thermal gas.

The density matrix of the thermal photon gas is given by \cite{Thermal_state},
\begin{equation}
	\hat{\rho}_{\rm Th} = \sum_{n=0}^{\infty} \frac{\langle \hat{n}\rangle^n}{(1+\langle \hat{n}\rangle)^{1+n}} \ket{n}\bra{n},
\end{equation}
where  $\langle n\rangle$ is the average number of photons. 
The corresponding Husimi distribution of thermal photons expressed in terms of the dimensionless classical variables $x,p$ takes the form of a symmetric Gaussian function, which can be written as,
\begin{eqnarray}
	Q_{\rm Th} = \frac{1}{\pi(1+\langle n\rangle)}\exp\left(-\frac{r^2}{2(1+\langle n\rangle)}\right),
	\label{Thermal_Husimi}
\end{eqnarray} 
where $r^2 = M(x^2+p^2)$.
For a comparison with the thermal state, we obtain the angular averaged Husimi function $\overline{Q}(r)=\int_{0}^{2\pi}Q(r,\theta)d\theta$ of the photon field after the quench dynamics. Although the spread in the phase space distribution is comparable to that of the thermal state, their detailed structure differs, which is illustrated in Fig.\ref{fig16}(a). Notably, a peak appears in the phase space distribution of the photon field near the ring of fixed points. Interestingly, such a scenario indicates the retention of memory of the underlying steady states, even when they become unstable after quench, which is similar to scarring phenomena \cite{Scar1,Scar2,Scar3,Scar4}. Consequently, the entanglement entropy $\mathcal{S}_L^p$ saturates to a slightly lower value compared to the entropy of the thermal state, as depicted in Fig.\ref{fig16}(b).  

The approach to a thermal state can be investigated experimentally through non-equilibrium dynamics of photon field in JCJJ by initially preparing identical coherent states in both cavities with a total energy corresponding to the $\pi$ mode. A subsequent quench to the unstable domain of $\pi$ mode can be performed by suitably choosing the strength of Kerr nonlinearity or photon hopping. 
Both the cavity and circuit QED can serve as ideal platforms to achieve such tunable parameters \cite{JCH_Plenio,Kerr_7}. Moreover, the optomechanical systems can also be used to generate  Kerr nonlinearity in a controlled manner \cite{Kerr_8}.
In this context, it is noteworthy that the thermalization of the photonic gas has already been observed in experimental setups such as optical microwave resonators \cite{Photonic_Thermalization1,Non_classical_2}. Apart from the coherence properties, such non-equilibrium dynamics of the photon fields can also reveal interesting phenomena, which can be probed in experiments \cite{Esslinger_SS_2,Hemmarich}.

\section{Conclusion}
\label{Conclusion}

To summarize, we explore the non-equilibrium dynamics of the Jaynes-Cummings dimer model in the presence of Kerr nonlinearity, focusing on the quantum states of photons as well as entanglement properties corresponding to the different dynamical states. 
Within the semiclassical approach, we systematically study the dynamics to chart out a variety of steady states and their regime of stability for different atom-photon coupling strengths and Kerr nonlinearity. Moreover, the stability analysis yields the frequency of photonic Josephson oscillation that can be probed in experiments.
The various branches of dynamical states are categorized based on the relative spin orientation and photon population imbalance between the cavities. Self trapped states with unequal photon populations in the two cavities emerge as a consequence of the different transitions. Apart from a perfect self trapped state, arising from a saddle-node bifurcation, we also identify two different self trapped states for which the Kerr interaction plays an important role. From quantum dynamics, we also observe the characteristic features of different steady states obtained semiclassically. However, interactions and atom-photon entanglement give rise to intriguing quantum effects, leading to a deviation from classicality. In contrast to the classical motion, dephasing is observed in spin dynamics as a result of relatively large quantum fluctuations in the spin $1/2$ qubits.
During the time evolution, the state of the photon field deviates from the initial coherent state and gradually loses its coherence due to phase fluctuation, which is typically enhanced by the Kerr nonlinearity. Apart from the phase fluctuations, we identify a periodic revival phenomenon for the self trapped state, exhibiting fascinating phase space structures of the photon field, particularly the appearance of a bimodal density distribution, resembling a cat state of the photon. Interestingly, photon mediated entanglement between the two atomic qubits, which are otherwise non interacting, makes JCJJ a promising candidate for quantum information processing. Using mutual information, we demonstrate how the quantum correlation between the atomic qubits in the two cavities can be manipulated by changing the photon population imbalance. Finally, we investigate the quench dynamics starting from a stable steady state to an unstable regime, which results in the formation of an incoherent gas of photons spread over the phase space, resembling a thermal state.

The Jaynes-Cummings dimer has already been realized in a circuit QED setup \cite{JC_dimer_expt}, as well it can be engineered by coupling the optical cavities \cite{JCH_Plenio,Plenio_Review1,Hopping_Houck}. The signature of the self-trapping phenomena has also been observed experimentally in micro-cavities \cite{Self_trapping1}, which is promising for the observation of different types of photonic Josephson oscillations discussed in this work. The Kerr nonlinearity can be realized in circuit QED \cite{Kerr_2,Kerr_7} as well as in optical cavities \cite{Kerr_3,Kerr_4,Kerr_5,Kerr_6}, which is the key ingredient for the observation of various quantum phenomena related to the steady states, such as the revival cycle in self trapped regimes.
A rich variety of collective phenomena can also be observed in cavities containing many atoms, which have been implemented in experiments by coupling the condensates of ultracold atoms with a cavity mode \cite{Esslinger_SS_2,Esslinger_SS_3,Hemmarich}.

Since dissipation is inherent in such atom-photon interacting systems,
they serve as an ideal platform to explore out of equilibrium dissipative phenomena \cite{Dissipative_transition3,Dissipative_transition4,Dissipative_transition5,Dissipative_transition6,Dissipative_transition7}, which has gained momentum  due to recent experiments \cite{Dissipative_transition1,Dissipative_transition2}.
Consequently, the dynamical states of JCJJ can acquire a finite lifetime, primarily due to weak dissipation arising from photon loss, which requires further investigation.

In conclusion, the Josephson coupled Jaynes-Cumming dimer can serve as a test bed to study the fascinating non-equilibrium phenomena, as well as manipulation of entanglement between the two atomic qubits, which can have potential applications in quantum information processing.
\begin{acknowledgments}
We thank Nirmalya Ghosh, Sudip Sinha and Sayak Ray for comments and fruitful discussions.
\end{acknowledgments}

\appendix
\section{Comparison between the cat state and dynamically generated bimodal distribution of photons}\label{Appendix_a}
\begin{figure}[b]
	\centering
	\includegraphics[width=\columnwidth]{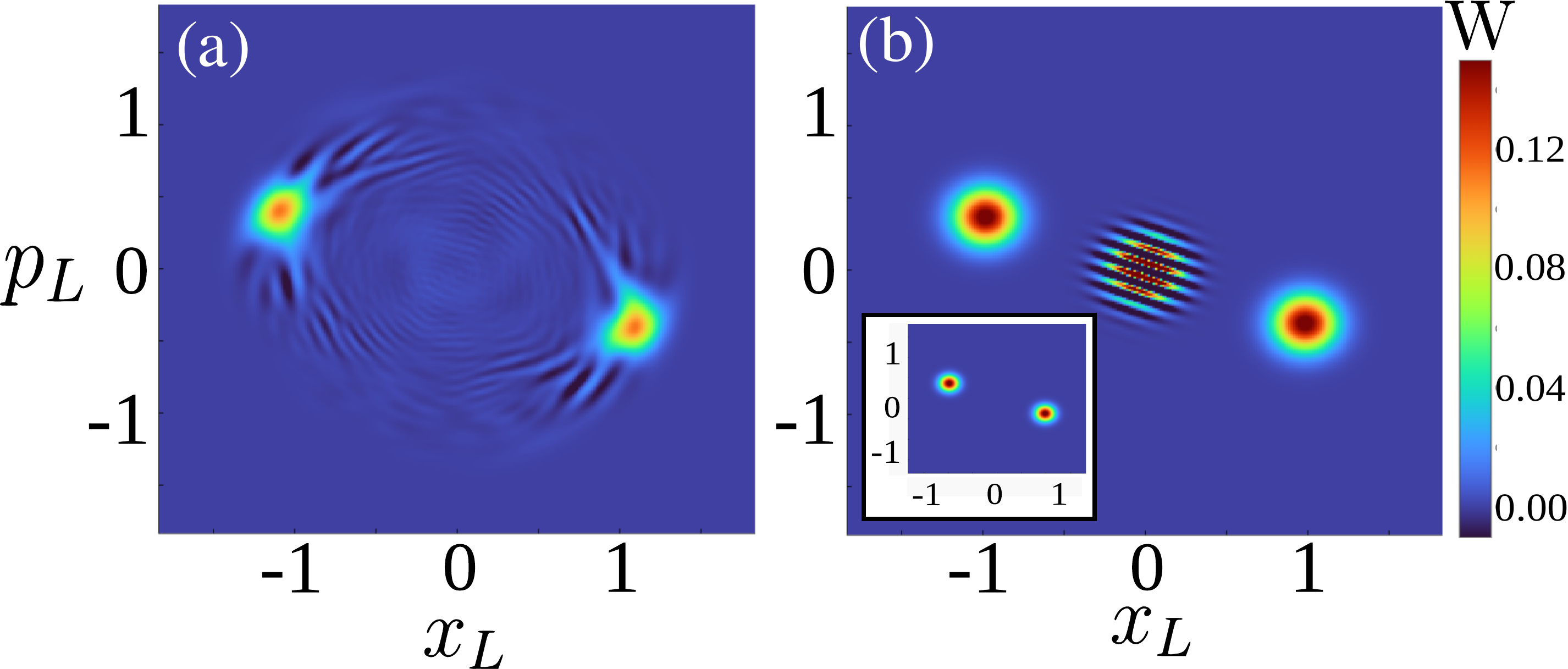}
	\caption{{\it Distinguishing the dynamical state of the photon field from the cat state in terms of the Wigner function:} The distribution (a) corresponds to the quantum state representing the photon field in the left cavity in the middle of the phase revival cycle (at a time marked by a square in Fig.\ref{fig14} of the main text) and (b) the cat state given in Eq.\eqref{cat_state}. The inset of (b) represents the Wigner function of the mixed density matrix $\hat{\rho}_L^m$ given in Eq.\eqref{Mixed_density_matrix}, which does not show any coherent oscillation between the two peaks, unlike the cat state. }
	\label{fig17}
\end{figure}

Here, we present a comparison between the photonic cat state and the bimodal phase space distribution of photons, which appears during the revival cycle of the self trapped (ST$_1$) state, shown in Fig.\ref{fig14}(c) of Sec.\ref{5}. The Husimi distribution corresponding to this photonic state in a particular cavity exhibits sharply peaked densities at the diametrically opposite points in phase space, which resembles the probability distribution of a cat state \cite{Dan_Walls,Cat_1,Cat_2,Cat_3,Cat_4,Cat_5} given by,
\begin{eqnarray}
	\ket{\psi_{\rm cat}} = \frac{\ket{\alpha}+\ket{-\alpha}}{\sqrt{2(1+e^{-2|\alpha|^2})}},
	\label{cat_state}
\end{eqnarray} 
where the density is peaked at the two phase space points $\alpha,-\alpha$.
Although the cat state is a superposition of two coherent states, their interference effect in the probability distribution almost disappears when they are far apart in phase space with $|\alpha|^2\gg 1$. On the other hand, the Wigner function of the cat state reveals the interference pattern at the center, as seen in Fig.\ref{fig17}(b). 

In contrast, the Wigner function of the density matrix representing the dynamical state of the photons in a particular cavity does not exhibit any interference pattern. Nevertheless, it is localized at two diagonally opposite points in the phase space, similar to a cat state (see Fig.\ref{fig17}(a)). Furthermore, the reduced density matrix of the photon field in each cavity represents a mixed state, as well as the phase fluctuation approaches close to its maximum limit (See Fig.\ref{fig14}(b)). Consequently, due to the lack of coherence, such a dynamical state of the photons can not be described by a simple linear combination of two coherent states similar to the cat state. Instead, this photonic state closely resembles an incoherent mixture of two coherent states represented by the density matrix,
\begin{eqnarray}
	\hat{\rho}_i^m &=& \frac{\ket{\alpha_i}\bra{\alpha_i}+\ket{-\alpha_i}\bra{-\alpha_i}}{2},
	\label{Mixed_density_matrix}
\end{eqnarray}
which exhibits the bimodal phase space distribution without any interference pattern (see inset of Fig.\ref{fig17}(b)).


\begin{thebibliography}{99}
	\bibitem{Angelakis}
	Changsuk Noh and Dimitris G. Angelakis, Rep. Prog. Phys. {\bf 80}, 016401 (2016).
	
	
	\bibitem{Serge_Haroche}
	J. M. Raimond, M. Brune, and S. Haroche, Rev. Mod. Phys. {\bf 73}, 565 (2001).
	
	\bibitem{Steven_Girvin}
	A. Blais, A. L. Grimsmo, S. M. Girvin, and A. Wallraff, Rev. Mod. Phys. {\bf 93}, 025005 (2021).
	
	\bibitem{QPT_0}
	K. Hepp and E. H. Lieb, Ann. Phys. (N.Y.) {\bf 76}, 360 (1973).
	
    \bibitem{QPT_Chaos_Brandes}
	C. Emary and T. Brandes, Phys. Rev. Lett. {\bf 90}, 044101 (2003);  Phys. Rev. E {\bf 67}, 066203 (2003).
	
	\bibitem{QPT_2}
	A. Baksic and C. Ciuti, Phys. Rev. Lett. {\bf 112}, 173601 (2014).
	
	
	\bibitem{Haake_2012}
	A. Altland and F. Haake, Phys. Rev. Lett. {\bf 108}, 073601 (2012); New J. Phys.
	{\bf 14}, 073011 (2012).
	
	\bibitem{Lea_2023}
	D. Villase\~{n}or, S. Pilatowsky-Cameo, M. A. Bastarrachea-Magnani, S. Lerma-Hern\'andez, L. F. Santos, and J. G. Hirsch, Entropy {\bf 25}, 8 (2022).
	
	\bibitem{Scar1}
	S. Pilatowsky-Cameo, D. Villase\~{n}or, M. A. Bastarrachea-Magnani, S. Lerma-Hern\'{a}ndez, L. F. Santos and J. G. Hirsch, Nat Commun {\bf 12}, 852 (2021).
	
	\bibitem{Scar2}
	S. Sinha, S. Ray, and S. Sinha, Journal of Physics: Condensed Matter {\bf 33}, 174005 (2021).
	
	
%
	
%
%
%

	
	\bibitem{Revival_1}
	J. Braum\"{u}ller, M. Marthaler, A. Schneider, A. Stehli, H. Rotzinger, M. Weides, and A. V. Ustinov, Nat Commun {\bf 8}, 779 (2017).

	\bibitem{Revival_2}
	J. I. Cirac, R. Blatt, A. S. Parkins, and P. Zoller, Phys. Rev. A {\bf 49}, 1202 (1994).

	\bibitem{Revival_3}
	A. De and R. Joynt, Phys. Rev. A {\bf 87}, 042336 (2013).
	
	\bibitem{Cat_1}
	L. Sun, A. Petrenko, Z. Leghtas, B. Vlastakis, G. Kirchmair, K. M. Sliwa, A. Narla, M. Hatridge, S. Shankar, J. Blumoff, L. Frunzio, M. Mirrahimi, M. H. Devoret and R. J. Schoelkopf, Nature {\bf 511}, 444–448 (2014).
	
	\bibitem{Cat_2}
	X. Pan, J. Schwinger, Ni-Ni Huang, P. Song, W. Chua, F. Hanamura, A. Joshi, F. Valadares, R. Filip, and Y. Y. Gao, Phys. Rev. X {\bf 13}, 021004 (2023).
	
	\bibitem{Cat_3}
	J. Lebreuilly, C. Aron, and C. Mora, Phys. Rev. Lett. {\bf 122}, 120402 (2019).
	
	\bibitem{Cat_4}
	S. M. Girvin, arXiv:1710.03179.
	
	\bibitem{Cat_5}
	V. Bu\v{z}ek, H. Moya-Cessa, P. L. Knight, and S. J. D. Phoenix, Phys. Rev. A {\bf 45}, 8190 (1992).
	
	\bibitem{Non_classical_1}
	C. Navarrete-Benlloch, J. J. Garc\'{i}a-Ripoll, and D. Porras, Phys. Rev. Lett. {\bf 113}, 193601 (2014).
	
	\bibitem{Non_classical_2}
	M. J. Gullans, J. Stehlik, Y.-Y. Liu, C. Eichler, J. R. Petta, and J. M. Taylor, Phys. Rev. Lett. {\bf 117}, 056801 (2016).
	
	\bibitem{Non_classical_3}
	G. Gasse, C. Lupien, and B. Reulet, Phys. Rev. Lett. {\bf 111}, 136601 (2013).
	
	\bibitem{Non_classical_4}
	S. J. Masson and S. Parkins, Phys. Rev. A {\bf 99}, 023822 (2019).
	
	\bibitem{Non_classical_5}
	A. Majumdar, M. Bajcsy, and J. Vu\v{c}kovi\'{c}, Phys. Rev. A {\bf 85}, 041801(R) (2012).
	
	\bibitem{Non_classical_6}
	K. Weiher, E. Agudelo, and M. Bohmann, Phys. Rev. A {\bf 100}, 043812 (2019).
	
	
	\bibitem{Esslinger_SS_2}
	J. L\'{e}onard, A. Morales, P. Zupancic, T. Esslinger, and T. Donner, Nature (London) {\bf 543}, 87 (2017).
	
	\bibitem{Esslinger_SS_3}
	J. L\'{e}onard, A. Morales, P. Zupancic, T. Donner, and T. Esslinger, Science {\bf 358}, 1415 (2017).
	
	
	\bibitem{Hemmarich}
	J. Klinder, H. Keßler, M. Wolke, L. Mathey, and A. Hemmerich, Proc. Natl. Acad. Sci. U.S.A. {\bf 112}, 3290 (2015).
	
	
	\bibitem{Dissipative_transition1}
	F. Brennecke, R. Mottl, K. Baumann, R. Landig, T. Donner, and T. Esslinger, Proc. Natl. Acad. Sci. USA {\bf 110}, 11763 (2013).
	
	\bibitem{Dissipative_transition2}
	M. Fitzpatrick, N. M. Sundaresan, A. C. Y. Li, J. Koch and A. A. Houck, Phys. Rev. X {\bf 7}, 011016 (2017).
	
	\bibitem{Dissipative_transition3}
	H. J. Carmichael, Phys. Rev. X {\bf 5}, 031028 (2015).
	
	\bibitem{Dissipative_transition4}
	F. Reiter, T. L. Nguyen, J. P. Home, and S. F. Yelin, Phys. Rev. Lett. {\bf 125}, 233602 (2020).
	
	
	\bibitem{Dissipative_transition5}
	K. C. Stitely, A. Giraldo, B. Krauskopf, and S. Parkins, Phys. Rev. Research {\bf 2}, 033131 (2020).
	
	
	\bibitem{Dissipative_transition6}
	J. Li, R. Fazio, and S. Chesi, New J. Phys. {\bf 24}, 083039 (2022).
	
	\bibitem{Dissipative_transition7}
	W. Kopylov, M. Radonji\'{c}, T. Brandes, A. Bala\v{z}, and A. Pelster, Phys. Rev. A {\bf 92}, 063832 (2015).
	
	
	%
	
	\bibitem{JC}
	E.T. Jaynes, F.W. Cummings, Proc. IEEE. {\bf 51} (1): 89–109 (1963).
	
	\bibitem{TC}
	M. Tavis and F. W. Cummings, Phys. Rev. {\bf 170}, 379 (1968).
	
	\bibitem{JCH_Plenio}
	M. Hartmann, F. G. S. L. Brand\~{a}o and M. B. Plenio, Nature Phys {\bf 2}, 849–855 (2006).
	
	\bibitem{Greentree}
	A. Greentree, C. Tahan, J. Cole et al., Nature Phys {\bf 2}, 856–861 (2006). 
	
	\bibitem{Plenio_Review1}
	G. Lepert, M. Trupke, M. J. Hartman, M. B. Plenio, and E. A. Hinds, New J. Phys. {\bf 13}, 113002 (2011).
	
	
	\bibitem{Hartmann}
	Michael J. Hartmann, J. Opt. {\bf 18}, 104005 (2016).
	
	\bibitem{Carusotto_Review}
	I. Carusotto and C. Ciuti, Rev. Mod. Phys. {\bf 85}, 299 (2013).
	
	\bibitem{Hopping_Houck}
	D. L. Underwood, W. E. Shanks, Jens Koch and A. A. Houck, Phys. Rev. A {\bf 86}, 023837 (2012).
	
	\bibitem{Superfulidity_light}
	P. Leboeuf and S. Moulieras, Phys. Rev. Lett. {\bf 105}, 163904 (2010).
	
	\bibitem{Vortex_Dominici}
	L. Dominici, R. Carretero-Gonz\'{a}lez, A. Gianfrate et al., Nat Commun {\bf 9}, 1467 (2018).
	
	\bibitem{Vortex_Carusotto}
	K. Lagoudakis, M. Wouters, M. Richard et al., Nature Phys {\bf 4}, 706–710 (2008).
	
	
	\bibitem{Plenio_effective_spin_system}
	M. J. Hartmann, F. G. S. L. Brand\~{a}o, and M. B. Plenio, Phys. Rev. Lett. {\bf 99}, 160501 (2007).
	
	\bibitem{Polariton_Yamamoto}
	T. Byrnes, N. Kim and Y. Yamamoto, Nature Phys {\bf 10}, 803–813 (2014).
	
	\bibitem{Fazio_glassy_phase}
	D. Rossini and R. Fazio, Phys. Rev. Lett. {\bf 99}, 186401 (2007).
	
	
	\bibitem{Hall_Sugato}
	J. Cho, D. G. Angelakis, and S. Bose, Phys. Rev. Lett. {\bf 101}, 246809 (2008).
	
	\bibitem{Girvin_time_reversal}
	J. Koch, A. A. Houck, K. Le Hur, and S. M. Girvin, Phys. Rev. A {\bf 82}, 043811 (2010).
	
	
	\bibitem{Blatter}
	S. Schmidt and G. Blatter, Phys. Rev. Lett. {\bf 103}, 086403 (2009).
	
	\bibitem{Le_Hur}
	J. Koch and K. Le Hur, Phys. Rev. A {\bf 80}, 023811 (2009).
	
	
	\bibitem{M_Knap}
	M. Knap, E. Arrigoni, and W. von der Linden, Phys. Rev. B {\bf 82}, 045126 (2010).
	
	
	\bibitem{Sibastian_1}
	L. Guo, S. Greschner, S. Zhu, and W. Zhang, Phys. Rev. A {\bf 100}, 033614 (2019).
	
	\bibitem{Sugato_bose}
	D. G. Angelakis, M. F. Santos, and S. Bose, Phys. Rev. A {\bf 76}, 031805(R) (2007).
	
	\bibitem{Yamamoto_Glass}
	N. Na, S. Utsunomiya, L. Tian, and Y. Yamamoto, Phys. Rev. A {\bf 77}, 031803 (2008).
	
	
	
	\bibitem{JC_dimer_expt}
	J. Raftery, D. Sadri, S. Schmidt, H. E. T\"{u}reci, and A. A. Houck, Phys. Rev. X {\bf 4},
	031043 (2014).
	
	\bibitem{JCD_Houck}
	S. Schmidt, D. Gerace, A. A. Houck, G. Blatter, and H. E. T\"{u}reci, Phys. Rev. B {\bf 82}, 100507(R) (2010).
	
	\bibitem{JCD_Manus_K}
	A. Dey and M. Kulkarni, Phys. Rev. A {\bf 101}, 043801 (2020).
	
		
	\bibitem{Dan_Walls}
	D. F. Walls and G. J. Milburn (1995).  {\it Quantum optics}.  Berlin; New York: Springer-Verlag.
	
	\bibitem{Kerr_1} 
	S. D. Du and C. D. Gong, Phys. Rev. A {\bf 50}, 779 (1994).
	
	
	\bibitem{Kerr_2}
	S. Rebi\'{c}, J.Twamley, and G. J. Milburn, Phys. Rev. Lett. {\bf 103}, 150503 (2009).
	
	\bibitem{Kerr_3}
	A. Imamo\u{g}lu, H. Schmidt, G. Woods, and M. Deutsch, Phys. Rev. Lett. {\bf 79}, 1467 (1997).
	
	\bibitem{Kerr_4}
	 H. Schmidt and A. Imamo\u{g}lu, Opt. Lett. {\bf 21}, 1936 (1996).
	
	\bibitem{Kerr_5}
	H. Rokhsari and K. J. Vahala, Opt. Lett. {\bf 30}, 427 (2005).
	
	\bibitem{Kerr_6}
	S. Rebic, S. M. Tan, A. S. Parkins, and D. F. Walls, J. Opt. B {\bf 1}, 490 (1999).
	
	\bibitem{Kerr_7}
	M. Kounalakis, C. Dickel, A. Bruno, N. Langford, and G. Steele, npj Quantum Inf. {\bf 4}, 38 (2018).
	
	\bibitem{Kerr_8}
	Z. R. Gong, H. Ian, Yu-xi Liu, C. P. Sun, and F. Nori, Phys. Rev. A {\bf 80}, 065801 (2009).
	
	\bibitem{JCD_Sadri}
	H. Shapourian and D. Sadri, Phys. Rev. A {\bf 93}, 013845 (2016).
	

	
	\bibitem{Dirac}
	P.A.M. Dirac, Proc. Cambridge Philos. Soc. {\bf 26}, 376 (1930); J. Frenkel, {\it Wave Mechanics}, Claredon Press, Oxford, 1934.
	
	\bibitem{Coherent_state}
	J. M. Radcliffe, J. Phys. A: Gen.Phys., {\bf 4}, 313 (1971).
	
	\bibitem{Dyn_transition1}
	J. A. Muniz, D. Barberena, R. J. Lewis-Swan, D. J. Young, J. R. K. Cline, A. M. Rey, and J. K. Thompson, Nature {\bf 580}, 602 (2020).
	
	\bibitem{Dyn_transition2}
	S. A. Weidinger, M. Heyl, A. Silva, and M. Knap, Phys. Rev. B {\bf 96}, 134313 (2017).
	
	
	
	\bibitem{Shenoy1}
	A. Smerzi, S. Fantoni, S. Giovanazzi, and S. R. Shenoy, Phys. Rev. Lett. {\bf 79}, 4950 (1997).
	
	\bibitem{Shenoy2}
	S. Raghavan, A. Smerzi, S. Fantoni, and S. R. Shenoy, Phys. Rev. A {\bf 59}, 620 (1999).
	
	\bibitem{vardi1}
	E. Boukobza, M. Chuchem, D. Cohen, and A. Vardi, Phys. Rev. Lett. {\bf 102}, 180403 (2009).
	
	\bibitem{vardi2}
	E. Boukobza, M. G. Moore, D. Cohen, and A. Vardi, Phys. Rev. Lett. {\bf 104}, 240402 (2010).
	
	\bibitem{BJJ_self_trapped}
	M. Albiez, R. Gati, J. F\"{o}lling, S. Hunsmann, M. Cristiani, and M. K. Oberthaler, Phys. Rev. Lett. {\bf 95}, 010402 (2005).
	
	\bibitem{Oberthalar}
	R. Gati and M. K. Oberthaler, J. Phys. B {\bf 40}, R61 (2007).
	
	\bibitem{KAM}
	M. Tabor, “{\it Chaos and Integrability in nonlinear Dynamics}: An Introduction,” Wiley-Interscience, USA, 1989.
	
	\bibitem{Mode_softening}
	R. Mottl, F. Brennecke, K. Baumann, R. Landig, T. Donner, and T. Esslinger, Science {\bf 336}, 1570 (2012).
	
	\bibitem{dephasing}
	S. Pramanik, S. Bandyopadhyay, and M. Cahay, Phys. Rev. B {\bf 68}, 075313 (2003).
	
	\bibitem{Dissipation1}
	V. Sevriuk, K. Y. Tan, E. Hyypp\"{a}, M. Silveri, M. Partanen,
	M. Jenei, S. Masuda, J. Goetz, V. Vesterinen, L. Gr\"{o}nberg,
	and M. M\"{o}tt\"{o}nen, Appl. Phys. Lett. {\bf 115}, 082601 (2019).
	
	\bibitem{Dissipation2}
	A. J. Fleisher, D. A. Long, Q. Liu, and J. T. Hodges, Phys. Rev. A {\bf 93}, 013833 (2016).
	
	\bibitem{Dissipation3}
	T. Laupr\^{e}tre, C. Proux, R. Ghosh, S. Schwartz, F. Goldfarb, and F. Bretenaker, Opt. Lett. {\bf 36}, 1551 (2011).
	
	\bibitem{Dissipation4}
	Y. Okawachi, B. Y. Kim, Y. Zhao, X. Ji, M. Lipson, and A. L. Gaeta, Optica {\bf 8}, 1458–1461 (2021).
	
	\bibitem{Self_trapping1}
	M. Abbarchi, A. Amo, V. G. Sala, D. D. Solnyshkov, H.Flayac, L. Ferrier, I. Sagnes, E. Galopin, A. Lema\^{i}tre, G. Malpuech, and J. Bloch, Nat. Phys. {\bf 9}, 275 (2013).
	
	\bibitem{phaseBarnett}
	D. T. Pegg and S. M. Barnett, Phys. Rev. A {\bf 39}, 1665 (1989).

	\bibitem{Nilson_Chuang}
	M. Nielsen and I. Chuang, {\it Quantum Computation and Quantum Information} (Cambridge University Press, Cambridge, England, 2000).
	
	\bibitem{MI1}
	D. P. DiVincenzo, M. Horodecki, D. W. Leung, J. A. Smolin, and B. M. Terhal, Phys. Rev. Lett. {\bf 92}, 067902 (2004).

	\bibitem{MI2}
	G. Adesso and A. Datta, Phys. Rev. Lett. {\bf 105}, 030501 (2010).

	\bibitem{MI3}
	X.-M. Lu, J. Ma, Z. Xi, and X. Wang, Phys. Rev. A {\bf 83}, 012327 (2011).
	
	\bibitem{MI4}
	 L. Henderson and V. Vedral, J. Phys. A: Math. Gen. {\bf 34}, 6899 (2001).

	\bibitem{MI5}
	P. Das, D. S. Bhakuni, and A. Sharma, Phys. Rev. A {\bf 107}, 043706 (2023).

	
	\bibitem{revival}
	M. Greiner, O. Mandel, T. W. H\"{a}nsch, I. Bloch, Nature {\bf 419}, 51–54 (2002).
	
	\bibitem{Thermal_state}
	C. Gerry and P. Knight, {\it Introductory Quantum Optics} (Cambridge University Press, Cambridge, 2004).
	
	\bibitem{Scar3}
		S. Sinha and S. Sinha, Phys. Rev. Lett. {\bf 125}, 134101 (2020).
		
	\bibitem{Scar4}
		D. Mondal, S. Sinha, S. Ray, J. Kroha, and S. Sinha, Phys. Rev. A {\bf 106}, 043321 (2022).
	
	\bibitem{Photonic_Thermalization1}
	J. Klaers, F. Vewinger and M. Weitz, Nature Phys {\bf 6}, 512–515 (2010).
	
	
	
	
	
	
\end{thebibliography}
\end{document}